\UseRawInputEncoding

\documentclass[preprint,review,12pt]{elsarticle}

\usepackage{amssymb}
\usepackage{amsthm}
\usepackage{amsfonts}
\usepackage{amsmath}%
\usepackage{graphicx}
\usepackage[dvips,final]{epsfig}
\usepackage{setspace}
\usepackage{lineno}
\usepackage{geometry}

\renewcommand{\mathbf}[1]{\mbox{\boldmath$#1$\unboldmath}}

\newcommand{\ddo}[0]{\scriptsize\mathbf{d} = \mathbf{0}\normalsize\mathbf{}}
\newcommand{\edo}[0]{\scriptsize\mathbf{\dot E} = \mathbf{0}\normalsize\mathbf{}}
\newcommand{\xvdo}[0]{\scriptsize\mathring{\mathbf{X}}_v = \mathbf{0}\normalsize\mathbf{}}
\newcommand{\lvo}[0]{\scriptsize\mathbf{l}_v^{\circ} = \mathbf{0}\normalsize\mathbf{}}
\journal{Computers \& Structures (10.1016/j.compstruc.2015.09.001)}

\begin{document}

\begin{frontmatter}



\title{Fully anisotropic finite strain viscoelasticity based on a reverse
multiplicative decomposition and logarithmic strains}


\author{Marcos Latorre}
\ead{m.latorre.ferrus@upm.es}
\author{Francisco Javier Mont\'{a}ns\corref{cor1}}
\ead{fco.montans@upm.es}
\cortext[cor1]{Corresponding author. Tel.:+34 637 908 304.}

\address{Escuela T\'{e}cnica Superior de Ingenier\'{\i}a Aeron\'{a}utica y del Espacio\\Universidad Polit\'{e}cnica de Madrid\\
Plaza Cardenal Cisneros, 3, 28040-Madrid, Spain}

\begin{abstract}

In this paper we present a novel formulation for phenomenological anisotropic
finite visco-hyperelasticity. The formulation is based on a multiplicative
decomposition of the equilibrated deformation gradient into nonequilibrated elastic and viscous
contributions. The proposal in this paper is a decomposition
reversed respect to that from Sidoroff allowing for
anisotropic viscous contributions. Independent anisotropic stored energies are
employed for equilibrated and non-equilibrated parts. The formulation uses
logarithmic strain measures in order to be teamed with spline-based
hyperelasticity. Some examples compare the results with formulations that use
the Sidoroff decomposition and also show the enhanced capabilities of the present model.
\end{abstract}


\end{frontmatter}

\section{Introduction}

Rubberlike materials and biological tissues are capable of sustaining large strains and
are frequently considered quasi-incompressible and hyperelastic in finite element analyses, see for example
\cite{Bathe,Ogdenbook,HolzapfelBook,Fungbook,KojicBathebook,SimoHughesBook,BonetWoodbook}. In the observed behavior of these
materials, specially in biological tissues, there is frequently a
relevant viscous component \cite{HolzapfelBook,Fungbook}. Hence,
visco-hyperelastic models are very important in both the engineering and
biomechanics fields.

Among the many types of formulations proposed for isochoric
viscoelasticity, two approaches stand out in finite element simulations.
The first one was advocated by Simo \cite{SimoHughesBook,Simo1997visco-damage} and successfully used by other researchers, see
\cite{HolzapfelBook,HolzapfelGasser,Holzapfel1996,PenaMartinezPena,PenaPenaDoblare}, among others. This
formulation is based on stress-like internal variables and allows for
anisotropic stored energies. However, this formulation is not adequate for
large deviations from thermodynamical equilibrium \cite{ReeseGovindjee,Haslach} (i.e. \emph{finite linear viscoelasticity}).
Furthermore, the instantaneous and relaxed stored energies are usually
proportional \cite{SimoHughesBook,Simo1997visco-damage}.

The second approach has been proposed by Reese and Govindjee
\cite{ReeseGovindjee} and used also in References \cite{HolmesLoughran} and
\cite{PericDettmer} among others. In this approach the Sidoroff multiplicative
decomposition \cite{Sidoroff} is employed and the stored energy is separated
into equilibrated and nonequilibrated parts following the framework introduced
by Lubliner \cite{Lubliner1985}. The main advantage of this formulation is
that it is valid for deformations away from thermodynamical equilibrium and
that distinct instantaneous and relaxed stored energies may be considered. As
a drawback, the phenomenological formulation is only valid for
isotropy, although anisotropic formulations are possible following these ideas
and modelling the microstructure \cite{NguyenJonesBoyce}.

Recently we have developed a formulation following the ideas from Reese and
Govindjee which is valid for anisotropic hyperelasticity and for deformations
arbitrarily away from thermodynamic equilibrium (i.e. \emph{finite nonlinear
viscoelasticity}) \cite{LatMonCM2015}. This formulation uses the Sidoroff
multiplicative decomposition of the total (equilibrated) deformation gradient
into non-equilibrated elastic and viscous parts. The equilibrated and non-equilibrated
stored energies are formulated in terms of logarithmic strains. These strain
measures are intuitive \cite{LatMonIJSS2014,Fiala,LatMonIJSS2015} and allow for simple formulations in large
strain elasto-plasticity \cite{EterovicBathe,CamineroMontansBathe,MontansBenitezCaminero}.
Furthermore, they are employed in spline-based hyperelasticity
\cite{SusBat2009,LatMonCAS2013,LatMonCM2014}. Spline-based hyperelasticity introduced by Sussman and Bathe permits the exact (in
practice) replication of experimental data and also facilitates the
interpretation of the material behavior \cite{LatMonCM2015} in
visco-hyperelasticity. Furthermore, it may be formulated as to preserve both
theoretical and numerical material symmetries consistency \cite{LatMonEJM2015}%
. However, the inconvenience of the formulation of Reference \cite{LatMonCM2015} based on Sidoroff's decomposition is that whereas the stored
energies may be anisotropic, the viscous component should arguably be isotropic. This is due to the intermediate configuration imposed by the Sidoroff multiplicative
decomposition. Hence, only one relaxation time can be considered as an
independent parameter (i.e. obtained from an experiment). The relaxation times
for the remaining components are given by the prescribed stored energies
\cite{LatMonCM2015}.

The purpose of this paper is to present a formulation for anisotropic
visco-hyperelasticity in which both the stored energies and the viscous
contribution are anisotropic. Therefore, in orthotropy up to six independent relaxation
times may be independently prescribed, i.e. obtained from six different
experiments as for the case of isochoric spline-based orthotropic equilibrated and
nonequilibrated stored energies \cite{LatMonCM2014}. The procedure employs a
reversed multiplicative decomposition from that used by Sidoroff. The
intermediate configuration from this decomposition allows for the formulation
of the stored energies using the same structural tensors and, hence,
facilitates the use of anisotropic viscosity tensors. The algorithm is
introduced using a special co-rotational formulation in order to facilitate a
parallelism with the formulation introduced in Reference \cite{LatMonCM2015}. As an
inconvenience of the present formulation when compared to the one presented in
\cite{LatMonCM2015}, the resulting non-equilibrated consistent tangent moduli tensor is
slightly non-symmetric for off-axis nonproportional loading. However, for the
numerical nonproportional examples presented in this paper typically only one
additional iteration is employed when using a symmetrized tensor. For the case of nonproportional off-axes loading,
the observed behavior is also slightly different due to the also different multiplicative decomposition employed. Therefore,
if the viscosity is considered isotropic, the formulation given in
\cite{LatMonCM2015} may be preferred, but for more general anisotropic
viscosities, the present formulation must be employed.

In this paper we focus mainly on the large strain formulation using the
reversed decomposition. For a detailed small strains motivation and for some
concepts used in the kinematics of the multiplicative decomposition, the
reader can refer to Reference \cite{LatMonCM2015}.

\section{Sidoroff's and Reverse multiplicative
decompositions\label{Section - original vs reversed}}%

\begin{figure}
[ptb]
\begin{center}
\includegraphics[width=0.95\textwidth]%
{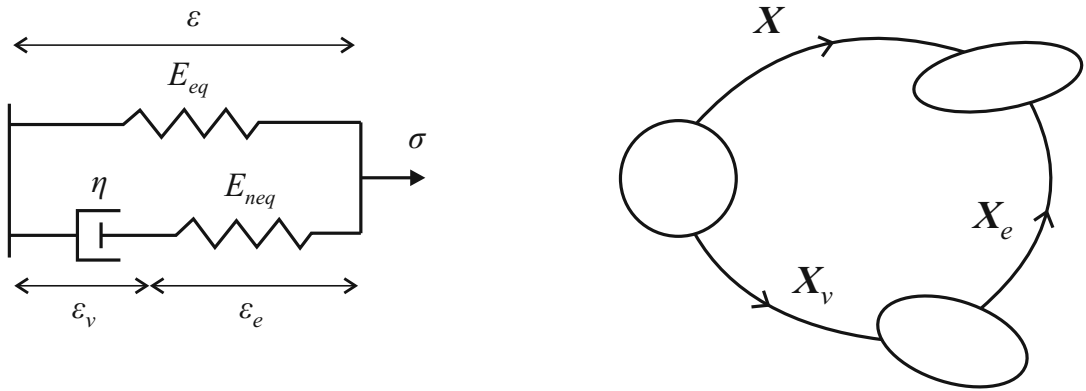}%
\caption{Sidoroff's multiplicative decomposition of the deformation gradient
$\mathbf{X}=\mathbf{X}_{e}\mathbf{X}_{v}$. Left: Equivalent standard linear solid. Right: Sidoroff multiplicative decomposition}%
\label{figure01.eps}%
\end{center}
\end{figure}

Unidimensional viscoelasticity is motivated by the standard solid rheological
model \cite{SimoHughesBook}, see Figure \ref{figure01.eps}, where the small
elongations of the springs and the viscous dashpot per unit device-length
(i.e. infinitesimal strains) are related through%
\begin{equation}
\varepsilon=\varepsilon_{e}+\varepsilon_{v}%
\end{equation}
Within the context of three-dimensional large deformations, a generalization
of this additive decomposition in terms of some finite deformation measure is
needed as point of departure in order to formulate strain-based constitutive
viscoelastic models. One possibility was proposed by Sidoroff \cite{Sidoroff}%
, who considered a multiplicative decomposition of the deformation gradient
motivated on the similar Lee multiplicative decomposition in elastoplasticity
\cite{Lee,Bilby} ---note that
this tensor is usually written as $\mathbf{F}$, but we adopt the notation
given in Ref. \cite{Bathe}
\begin{equation}
\mathbf{X}=\mathbf{X}_{e}\mathbf{X}_{v}
\label{Sidoroff's decomposition}%
\end{equation}
where $\mathbf{X}_{v}$ includes the viscous contribution to the
total deformation from the reference state to time $t$ and $\mathbf{X}_{e}$ accounts for the remaining elastic (nonequilibrated) contribution, see
Figure \ref{figure01.eps}. Motivated on the standard solid of Figure
\ref{figure01.eps}, the intermediate state may be interpreted as the internal, non-equilibrated
``stress-free'' configuration obtained by the virtual elastic unloading of the
\emph{equivalent} Maxwell element from the current configuration by means of
$\mathbf{X}_{e}^{-1}$ \cite{BergstromBoyce}. The hypothetically
relaxed total gradient is $\mathbf{X}^{\ast}=%
\mathbf{X}_{e}^{-1}\mathbf{X}=\mathbf{X}_{v}$. However, as
a clear difference with finite elastoplasticity, note that this internal
unloading is only fictitious even under homogeneous deformations (i.e. the intermediate configuration is not strictly speaking a ``stress-free'' configuration). The presence of an elastic deformation gradient
makes the system to be internally unbalanced, so the system is continuously
evolving in order to reach thermodynamic equilibrium in the sense that
$\mathbf{X}_{e}\rightarrow\mathbf{I}$ and $\mathbf{X}%
_{v}\rightarrow\mathbf{X}$, for a given (fixed) total gradient
$\mathbf{X}$. Hence, the intermediate configuration is truly relaxed
only when internal static equilibrium is attained. In that case, note that the
intermediate configuration has relaxed to (is coincident to) the actual
configuration, i.e. $\mathbf{X}=\mathbf{I}\mathbf{X}_{v}$
at $t\rightarrow\infty$.%

\begin{figure}
[ptb]
\begin{center}
\includegraphics[width=0.95\textwidth]%
{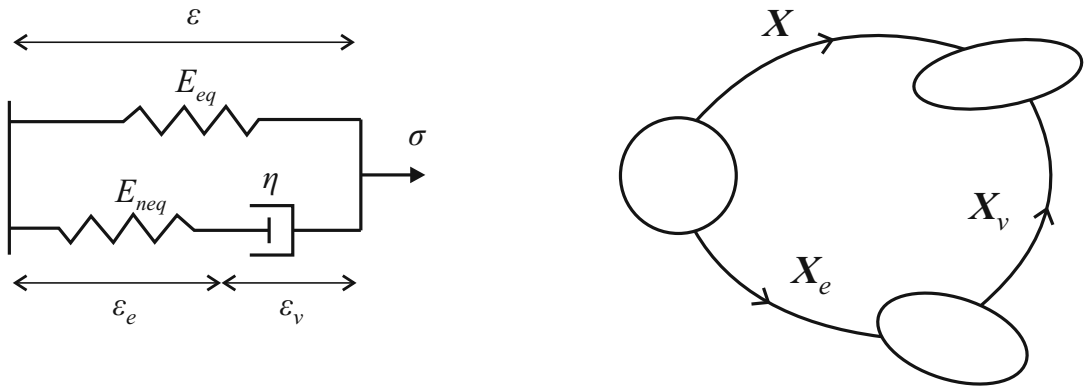}%
\caption{Reverse multiplicative decomposition of the deformation gradient
$\mathbf{X}=\mathbf{X}_{v}\mathbf{X}_{e}$. Left: Equivalent standard linear solid. Right: Reverse multiplicative decomposition}%
\label{figure02.eps}%
\end{center}
\end{figure}

Consider now the standard solid of Figure \ref{figure02.eps}. Of course, due
to the additive decomposition used in infinitesimal viscoelasticity, the
mechanical devices of Figures \ref{figure01.eps} and \ref{figure02.eps}
may be considered equivalent from a \emph{quantitative} standpoint. That is,
the same model for the small strains unidimensional case \cite{LatMonCM2015} is obtained in both cases.
However, they admit different physical interpretations
even for the small strains case. Furthermore, they lead to different
formulations in the large strains setting. The extension of
the rheological model shown in Figure \ref{figure02.eps} to the finite
deformation context is given by the \emph{reversed} multiplicative
decomposition of the deformation gradient%
\begin{equation}
\mathbf{X}=\mathbf{X}_{v}\mathbf{X}_{e}
\label{reversed Sidoroff's decomposition}%
\end{equation}
where $\mathbf{X}_{e}$ includes the elastic contribution to the
total deformation from the reference state to time $t$ and
$\mathbf{X}_{v}$ accounts for the remaining viscous contribution. An
apparent difference with the multiplicative decomposition of Figure
\ref{figure01.eps} is that in this second case, the virtual elastic
unloading of the \emph{equivalent} Maxwell element is performed from the
intermediate configuration to the reference configuration by means of
$\mathbf{X}_{e}^{-1}$. The same hypothetically relaxed total
gradient is obtained $\mathbf{X}^{\ast}=(\mathbf{X}%
_{v}\mathbf{X}_{e}^{-1}\mathbf{X}_{v}^{-1})%
\mathbf{X}=\mathbf{X}_{v}$ in this case (we emphasize that these situations
are only fictitious). However, for a fixed total gradient $%
\mathbf{X}$ the intermediate configuration will have
``relaxed'' to (will be coincident to) the reference configuration, i.e.
$\mathbf{X}=\mathbf{X}_{v}\mathbf{I}$ at $t\rightarrow
\infty$, which become the main difference between both
multiplicative decompositions.%

Viscoelasticity formulations based on strain-like internal variables are built
on the hypothesis of the existence of a strain energy density containing an
equilibrated contribution and a non-equilibrated one \cite{Lubliner1985,ReeseGovindjee}.
Given a multiplicative decomposition of the deformation
gradient $\mathbf{X}$ into an elastic part $\mathbf{X}_{e}$ and a viscous one
$\mathbf{X}_{v}$, the total stored energy function $\Psi$ is therefore written
as
\begin{equation}
\Psi=\Psi_{eq}\left(  \mathbf{A}\right)  +\Psi_{neq}\left(  \mathbf{A}%
_{e}\right)  \label{strain energy function A}%
\end{equation}
where $\mathbf{A}$ and $\mathbf{A}_{e}$ are the Green-Lagrange strain tensors
obtained from the total deformation gradient $\mathbf{X}$ and the internal
elastic gradient $\mathbf{X}_{e}$, respectively. Of course, $\Psi_{eq}$ and
$\Psi_{neq}$ may be expressed in terms of other Lagrangian strain measures.
However, we employ for now quadratic strain measures because they facilitate the
analytical derivation of the material formulation, as we show in the following section.

There exists a crucial difference in the non-equilibrated part of the
constitutive hypothesis given in Eq.(\ref{strain energy function A}) when one
uses either the Sidoroff or the reverse multiplicative decomposition of the
deformation gradient $\mathbf{X}$. On the one hand, if $\mathbf{X}%
=\mathbf{X}_{e}\mathbf{X}_{v}$, the non-equilibrated Green-Lagrange strains
$\mathbf{A}_{e}$ and the strain energy function $\Psi_{neq}$ are both defined in
the intermediate configuration. Hence, the second Piola-Kirchhoff stresses
that directly derive from $\Psi_{neq}$ (i.e. $\mathbf{S}_{neq}^{|e}=d\Psi_{neq}%
/d\mathbf{A}_{e}$, using
the notation introduced in Ref. \cite{LatMonCM2015}) also operates in that
configuration. On the other hand, if $\mathbf{X}=\mathbf{X}_{v}\mathbf{X}_{e}%
$, both $\mathbf{A}_{e}$ and $\Psi_{neq}$ are defined in the reference
configuration and $\mathbf{S}_{neq}^{|e}$ operates in the reference
configuration as well. This consideration will show relevant when deriving the
constitutive equation for the viscous flow in Section
\ref{Section - Constitutive equation for the viscous flow}.

\section{Finite strain viscoelasticity based on the reversed
decomposition\label{Section - Finite visco-hyperelasticity based on the reversed Sidoroff decomposition}%
}

In Ref. \cite{LatMonCM2015} we derived a computational model for finite fully
non-linear anisotropic visco-hyperelasticity based on the Sidoroff's
multiplicative decomposition of the deformation gradient given
in Eq. (\ref{Sidoroff's decomposition}). Departing from this kinematical
hypothesis, we show that the material formulation of the finite theory can be
derived following analogous steps to those followed in the infinitesimal case.
This is possible due to the fact that the second-order tensor on which depends
$\Psi_{neq}$ in Eq. (\ref{strain energy function A}), i.e. the
non-equilibrated elastic strain tensor $\mathbf{A}_{e}$, can be expressed as
an explicit function of $\mathbf{A}$ and $\mathbf{X}_{v}$, which may both be
taken as the independent variables of the continuum formulation%
\begin{equation}
\mathbf{A}_{e}\left(  \mathbf{A},\mathbf{X}_{v}\right)  =\mathbf{X}_{v}%
^{-T}\left(  \mathbf{A}-\mathbf{A}_{v}\right)  \mathbf{X}_{v}^{-1}%
=\mathbf{X}_{v}^{-T}\odot\mathbf{X}_{v}^{-T}:\left(  \mathbf{A}-\mathbf{A}%
_{v}\right)  \label{Ae multiplicative decomposition}%
\end{equation}
The symbol $\odot$ in the preceding expression denotes the mixed dyadic product
between second-order tensors $\left(  \mathbf{Y}\odot\mathbf{Z}\right)
_{ijkl}=Y_{ik}Z_{jl}$.

The setting is different for the reversed decomposition given in Eq.
(\ref{reversed Sidoroff's decomposition}) because an explicit expression of
$\mathbf{A}_{e}$ in terms of $\mathbf{A}$ and $\mathbf{X}_{v}$ is not found,
so the material formulation cannot be, a priori, derived as made with the
original multiplicative decomposition. However, we show next that the material
formulation for the reversed decomposition (and analogously, for the
Sidoroff's one)\ can be alternatively derived departing from the spatial
formulation. Subsequently, we will be able to develop a model for finite
anisotropic visco-hyperelasticity based on the reverse multiplicative
decomposition and logarithmic strains following similar conceptual steps to
those explained in Ref. \cite{LatMonCM2015}.

\subsection{Spatial description}

From the reverse multiplicative decomposition of the deformation gradient
$\mathbf{X}=\mathbf{X}_{v}\mathbf{X}_{e}$, the expression of the spatial
velocity gradient $\mathbf{l}=\mathbf{\dot{X}X}^{-1}$ in terms of the viscous
velocity gradient $\mathbf{l}_{v}=\mathbf{\dot{X}}_{v}\mathbf{X}_{v}^{-1}$ and
the elastic velocity gradient $\mathbf{l}_{e}=\mathbf{\dot{X}}_{e}%
\mathbf{X}_{e}^{-1}$ reads%
\begin{equation}
\mathbf{l}=\mathbf{l}_{v}+\mathbf{X}_{v}\mathbf{l}_{e}\mathbf{X}_{v}^{-1}
\label{velocity gradients}%
\end{equation}
where $\mathbf{l}$ and $\mathbf{l}_{v}$ operate in the current configuration
and $\mathbf{l}_{e}$ does in the intermediate configuration. As made in Ref.
\cite{LatMonCM2015}, we are interested herein in obtaining the elastic
deformation rate tensor $\mathbf{d}_{e}=sym(\mathbf{l}_{e})$ as a function of
both the deformation rate tensor $\mathbf{d}=sym(\mathbf{l})$ and the viscous
velocity gradient $\mathbf{l}_{v}$ (or some quantity related to the latter
one). From Eq. (\ref{velocity gradients}) we obtain in the intermediate configuration
\begin{equation}
\mathbf{d}_{e}\left(  \mathbf{d},\mathbf{l}_{v}^{\circ}\right)  =sym\left(
\mathbf{X}_{v}^{-1}\mathbf{dX}_{v}\right)  -sym\left(  \mathbf{X}_{v}%
^{-1}\mathbf{l}_{v}^{\circ}\mathbf{X}_{v}\right)  \label{deformation rates}%
\end{equation}
where $\mathbf{l}_{v}^{\circ}$ stands for the co-rotational viscous velocity
gradient%
\begin{equation}
\mathbf{l}_{v}^{\circ}:=\mathbf{l}_{v}-skew(\mathbf{l})=\mathbf{l}%
_{v}-\mathbf{w}=\left(
\mathbf{\dot{X}}_{v}-\mathbf{wX}_{v}\right)  \mathbf{X}_{v}^{-1}=:\mathbf{\mathring{X}}_{v}\mathbf{X}_{v}^{-1}
\label{co-rotational lv}%
\end{equation}
and $\mathbf{\mathring{X}}_{v}$ for the co-rotational rate of the viscous gradient.
Since the co-rotational total velocity gradient is given by $\mathbf{l}%
^{\circ}:=\mathbf{l}-\mathbf{w}=\mathbf{d}$, note that the two independent
variables in rate form in Eq. (\ref{deformation rates}) may be seen as
co-rotational (Jaumann-Zaremba) rates, i.e. $\mathbf{d}_{e}\left(
\mathbf{d},\mathbf{l}_{v}^{\circ}\right)  =\mathbf{d}_{e}\left(
\mathbf{l}^{\circ},\mathbf{l}_{v}^{\circ}\right)  $. Equation
(\ref{deformation rates}) may be rewritten as%
\begin{equation}
\mathbf{d}_{e}\left(  \mathbf{d},\mathbf{l}_{v}^{\circ}\right)  =\left.
\mathbb{M}_{d}^{d_{e}}\right\vert _{\lvo}:\mathbf{d}+\left.  \mathbb{M}%
_{l_{v}^{\circ}}^{d_{e}}\right\vert _{\ddo}:\mathbf{l}_{v}^{\circ}
\label{deformation rates 2}%
\end{equation}
where, for further use, we just identify the fourth-order
mapping tensor ---we omit (minor) symmetrization issues for the matter of
notation simplicity%
\begin{equation}
\left.  \mathbb{M}_{d}^{d_{e}}\right\vert _{\lvo}=\dfrac{1}{2}\left(
\mathbf{X}_{v}^{-1}\odot\mathbf{X}_{v}^{T}+\mathbf{X}_{v}^{T}\odot
\mathbf{X}_{v}^{-1}\right)  =:\mathbf{X}_{v}^{-1}\overset{s}{\odot}%
\mathbf{X}_{v}^{T} \label{mapping tensor d de}%
\end{equation}
This purely geometrical tensor lacks major symmetry in general and is responsible
for the lack of symmetry of the global tangent as it will be seen below. Equations
(\ref{deformation rates}) or (\ref{deformation rates 2}) may also be
interpreted as%
\begin{equation}
\mathbf{d}_{e}=\mathbf{d}_{e}\left(  \mathbf{d},\mathbf{0}\right)
+\mathbf{d}_{e}\left(  \mathbf{0},\mathbf{l}_{v}^{\circ}\right)  =\left.
\mathbf{d}_{e}\right\vert _{\lvo}+\left.  \mathbf{d}_{e}\right\vert _{\ddo}
\label{deformation rates de}%
\end{equation}
This interpretation will be useful in the two-step predictor-corrector
integration scheme used below. Note that with the consideration of the
co-rotational rate of the viscous gradient instead of its total rate, we
arrive at a formulation which is conceptually analogous to the spatial
formulation based on the Sidoroff's decomposition of Ref. \cite{LatMonCM2015}.
In this case the virtual state for which $\mathbf{l}_{v}^{\circ}=\mathbf{0}$,
representing the state in which the viscous velocity gradient relative to a
reference frame with spin $\mathbf{w}$ vanishes, naturally emerges from the
(reversed) kinematic decomposition given in Eq. (\ref{deformation rates}).
This state will define the internal kinematic constraint for no dissipation,
as we see below.

The material rate of Eq. (\ref{strain energy function A}) yields ---we use $d(\cdot)/d(\ast)$ to denote total differentiation
of $(\cdot)$ with respect to the tensorial variable $(\ast)$%
\begin{align}
\dot{\Psi}  &  =\dot{\Psi}_{eq}\left(  \mathbf{A}\right)  +\dot{\Psi}%
_{neq}\left(  \mathbf{A}_{e}\right) \nonumber\\
&  =\dfrac{d\Psi_{eq}}{d\mathbf{A}}:\mathbf{\dot{A}}%
+\dfrac{d\Psi_{neq}}{d\mathbf{A}_{e}}:\mathbf{\dot{A}}%
_{e}\nonumber\\
&  =\mathbf{S}_{eq}:\mathbf{\dot{A}}+\mathbf{S}_{neq}^{|e}:\mathbf{\dot{A}%
}_{e} \label{rate of strain energy A}%
\end{align}
where the superscript in expressions of the type $(\bullet)^{|e}$
indicates that the variable $(\bullet)$ has been obtained through differentiation with respect to
the internal \emph{elastic} strains ($\mathbf{A}_{e}$ in this case). This distinction will show
relevant below. Since $\mathbf{\dot{A}}_{e}$ is the pull-back of
$\mathbf{d}_{e}$ from the intermediate configuration to the reference
configuration by means of%
\begin{equation}
\mathbf{\dot{A}}_{e}=\mathbf{X}_{e}^{T}\mathbf{d}_{e}\mathbf{X}_{e}%
=\mathbf{X}_{e}^{T}\odot\mathbf{X}_{e}^{T}:\mathbf{d}_{e}=:\mathbb{M}_{d_{e}%
}^{\dot{A}_{e}}:\mathbf{d}_{e} \label{Aedot de}%
\end{equation}
and $\mathbf{\dot{A}}$ is the pull-back of $\mathbf{d}$ from the actual
configuration to the reference configuration%
\begin{equation}
\mathbf{\dot{A}}=\mathbf{X}^{T}\mathbf{dX}=\mathbf{X}^{T}\odot\mathbf{X}%
^{T}:\mathbf{d}=:\mathbb{M}_{d}^{\dot{A}}:\mathbf{d} \label{Adot d}%
\end{equation}
the respective push-forward operations of the terms in the right-hand side of
Eq. (\ref{rate of strain energy A}) become%
\begin{align}
\dot{\Psi}  &  =\mathbf{S}_{eq}:\mathbf{X}^{T}\mathbf{dX}+\mathbf{S%
}_{neq}^{|e}:\mathbf{X}_{e}^{T}\mathbf{d}_{e}\mathbf{X}_{e}\nonumber\\
&  =\mathbf{XS}_{eq}\mathbf{X}^{T}:\mathbf{d}+\mathbf{X}_{e}\mathbf{S%
}_{neq}^{|e}\mathbf{X}_{e}^{T}:\mathbf{d}_{e}\nonumber\\
&  =\mathbf{\tau}_{eq}:\mathbf{d}+\mathbf{\tau}_{neq}^{|e}:\mathbf{d}_{e}
\label{rate of strain energy d}%
\end{align}
In Eq. (\ref{rate of strain energy d}) we have defined the symmetric Kirchhoff
stress tensors $\mathbf{\tau}_{eq}$ (operating in the actual configuration)
and $\mathbf{\tau}_{neq}^{|e}$ (operating in the intermediate configuration)
as%
\begin{align}
\mathbf{\tau}_{eq}  &  :=\mathbf{XS}_{eq}\mathbf{X}^{T}=\mathbf{S}%
_{eq}:\mathbf{X}^{T}\odot\mathbf{X}^{T}=\mathbf{S}_{eq}:\mathbb{M}_{d}%
^{\dot{A}}\\
\mathbf{\tau}_{neq}^{|e}  &  :=\mathbf{X}_{e}\mathbf{S}_{neq}^{|e}%
\mathbf{X}_{e}^{T}=\mathbf{S}_{neq}^{|e}:\mathbf{X}_{e}^{T}\odot
\mathbf{X}_{e}^{T}=\mathbf{S}_{neq}^{|e}:\mathbb{M}_{d_{e}}^{\dot{A}_{e}}%
\end{align}
The insertion of Eq. (\ref{deformation rates 2}) into Eq.
(\ref{rate of strain energy d}) gives%
\begin{equation}
\dot{\Psi}=\underset{%
\begin{array}
[c]{c}%
\left.  \dot{\Psi}\right\vert _{\lvo}%
\end{array}
}{~\underbrace{\left(  \mathbf{\tau}_{eq}+\mathbf{\tau}_{neq}^{|e}:\left.
\mathbb{M}_{d}^{d_{e}}\right\vert _{\lvo}\right)  :\mathbf{d}}~}\underset{%
\begin{array}
[c]{c}%
\left.  \dot{\Psi}\right\vert _{\ddo}%
\end{array}
}{+~\underbrace{\mathbf{\tau}_{neq}^{|e}:\left.  \mathbb{M}_{l_{v}^{\circ}%
}^{d_{e}}\right\vert _{\ddo}:\mathbf{l}_{v}^{\circ}}~}%
\end{equation}
where the mapping tensors $\left.  \mathbb{M}_{d}^{d_{e}}\right\vert _{\lvo}$
and $\left.  \mathbb{M}_{l_{v}^{\circ}}^{d_{e}}\right\vert _{\ddo}$ perform
the adequate transformations to the spatial configuration for work-conjugacy.

The dissipation inequality in spatial description%
\begin{equation}
\left.  \mathbf{\tau}:\mathbf{d}-\dot{\Psi}\right.  =\left(  \mathbf{\tau
}-\mathbf{\tau}_{eq}-\mathbf{\tau}_{neq}^{|e}:\left.  \mathbb{M}_{d}^{d_{e}%
}\right\vert _{\lvo}\right)  :\mathbf{d}-\mathbf{\tau}_{neq}^{|e}:\left.
\mathbb{M}_{l_{v}^{\circ}}^{d_{e}}\right\vert _{\ddo}:\mathbf{l}_{v}^{\circ
}\geq0
\end{equation}
is fulfilled in any case if, first ($\mathbf{l}_{v}^{\circ}=\mathbf{0}$
implies no dissipation, so the equality must hold)%
\begin{equation}
\mathbf{\tau}=\mathbf{\tau}_{eq}+\mathbf{\tau}_{neq}^{|e}:\left.
\mathbb{M}_{d}^{d_{e}}\right\vert _{\lvo}=\mathbf{\tau}_{eq}+\mathbf{\tau
}_{neq} \label{tau stresses}%
\end{equation}
and, second, the Kirchhoff stresses $\mathbf{\tau}_{neq}^{|e}$\ dissipate
power with the pull-back of $\mathbf{l}_{v}^{\circ}$ from the current
configuration to the intermediate one%
\begin{equation}
-\mathbf{\tau}_{neq}^{|e}:\left.  \mathbb{M}_{l_{v}^{\circ}}^{d_{e}%
}\right\vert _{\ddo}:\mathbf{l}_{v}^{\circ}=\mathbf{\tau}_{neq}^{|e}%
:\mathbf{X}_{v}^{-1}\mathbf{l}_{v}^{\circ}\mathbf{X}_{v}\geq0
\label{dissipation inequality spatial 0}%
\end{equation}
where Eq. (\ref{deformation rates}) and the symmetry of $\mathbf{\tau%
}_{neq}^{|e}$ have been used. Equation (\ref{tau stresses}) shows that the existing
geometrical mapping between the non-equilibrated Kirchhoff stress tensors
$\mathbf{\tau}_{neq}$, operating in the actual configuration, and
$\mathbf{\tau}_{neq}^{|e}$, defined in the intermediate configuration, is
given by the same mapping tensor that relates $\mathbf{d}$ to $\mathbf{d}_{e}$
when $\mathbf{l}_{v}^{\circ}=\mathbf{0}$, i.e that of Eq.
(\ref{mapping tensor d de}) ---compare to the original Sidoroff's
decomposition where $\mathbf{\tau}_{neq}=\mathbf{\tau}_{neq}^{|e}$%
\begin{align}
\mathbf{\tau}_{neq}=\mathbf{\tau}_{neq}^{|e}:\left.  \mathbb{M}_{d}^{d_{e}%
}\right\vert _{\lvo}&=\mathbf{\tau}_{neq}^{|e}:\mathbf{X}_{v}^{-1}\overset
{s}{\odot}\mathbf{X}_{v}^{T}\nonumber\\
&=\dfrac{1}{2}\left(  \mathbf{X}_{v}^{-T}%
\mathbf{\tau}_{neq}^{|e}\mathbf{X}_{v}^{T}+\mathbf{X}_{v}\mathbf{\tau%
}_{neq}^{|e}\mathbf{X}_{v}^{-1}\right)  \label{tau stresses neq}%
\end{align}
From the definition of $\mathbf{\tau}_{neq}$ in terms of $\mathbf{\tau%
}_{neq}^{|e}$, we notice the equivalence between the following non-dissipative
mechanical powers%
\begin{equation}
\mathbf{\tau}_{neq}:\mathbf{d}=\mathbf{\tau}_{neq}^{|e}:\left.
\mathbf{d}_{e}\right\vert _{\lvo}=\left.  \dot{\Psi}_{neq}\right\vert _{\lvo}
\label{non-dissipative energy rate}%
\end{equation}
Finally, the dissipated power due to viscous effects given in Eq.
(\ref{dissipation inequality spatial 0}) can be rewritten using Eq.
(\ref{deformation rates de}) as%
\begin{equation}
-\mathbf{\tau}_{neq}^{|e}:\left.  \mathbb{M}_{l_{v}^{\circ}}^{d_{e}%
}\right\vert _{\ddo}:\mathbf{l}_{v}^{\circ}=-\mathbf{\tau}_{neq}^{|e}:\left.
\mathbf{d}_{e}\right\vert _{\ddo}\geq0 \label{dissipation inequality spatial}%
\end{equation}
which can be read as%
\begin{equation}
\left.  \dot{\Psi}_{neq}\right\vert _{\ddo}\leq0
\end{equation}
The dissipation inequality in spatial description given in Eq.
(\ref{dissipation inequality spatial}) will let us define a general
anisotropic constitutive equation for the viscous flow based on material
elastic logarithmic strains in the next sections.

\subsection{Material description\label{Section - Material description}}

From Eqs. (\ref{deformation rates 2})--(\ref{deformation rates de}) we obtain%
\begin{equation}
\left.  \mathbf{d}_{e}\right\vert _{\lvo}=\left.  \mathbb{M}_{d}^{d_{e}%
}\right\vert _{\lvo}:\mathbf{d} \label{de d}%
\end{equation}
Using Eqs. (\ref{Aedot de}) and (\ref{Adot d}), the Lagrangian counterpart of
Eq. (\ref{de d}) is ---note that we can equivalently use the subscripts
$\mathbf{l}_{v}^{\circ}=\mathbf{0}$ or $\mathbf{\mathring{X}}_{v}=\mathbf{0}$
in order to refer to the same non-dissipative state%
\begin{equation}
\left.  \mathbf{\dot{A}}_{e}\right\vert _{\xvdo}=\mathbb{M}_{d_{e}}^{\dot
{A}_{e}}:\left.  \mathbb{M}_{d}^{d_{e}}\right\vert _{\lvo}:\mathbb{M}_{\dot
{A}}^{d}:\mathbf{\dot{A}}=\left.  \dfrac{\delta\mathbf{A}_{e}}{\delta
\mathbf{A}}\right\vert _{\xvdo}:\mathbf{\dot{A}} \label{Aedot Adot}%
\end{equation}
where we define the \emph{modified} partial gradient%
\begin{align}
\left.  \dfrac{\delta\mathbf{A}_{e}}{\delta\mathbf{A}}\right\vert _{\xvdo}  &
:=\mathbb{M}_{d_{e}}^{\dot{A}_{e}}:\left.  \mathbb{M}_{d}^{d_{e}}\right\vert
_{\lvo}:\mathbb{M}_{\dot{A}}^{d}\nonumber\\
&  =\mathbf{X}_{e}^{T}\odot\mathbf{X}_{e}^{T}:\mathbf{X}_{v}^{-1}\overset
{s}{\odot}\mathbf{X}_{v}^{T}:\mathbf{X}^{-T}\odot\mathbf{X}^{-T}\nonumber\\
&  =\mathbf{C}_{e}\mathbf{C}^{-1}\overset{s}{\odot}\mathbf{I}
\label{mapping tensor Ae A}%
\end{align}
as the fourth-order tensor that maps the strain rate $\mathbf{\dot{A}}$ to the
strain rate $\mathbf{\dot{A}}_{e}$ when there is no dissipation.
The same result given in Eq. (\ref{mapping tensor Ae A}) is obtained taking the time
derivative of the elastic right Cauchy-Green deformation tensor $\mathbf{C}_{e}$ (given in terms of the deformation gradient
$\mathbf{X}$ and the left Cauchy-Green deformation tensor $\mathbf{B}_{v}%
^{-1}=\mathbf{X}_{v}^{-T}\mathbf{X}_{v}^{-1}$ as $\mathbf{C}_{e}=\mathbf{X}^{T}\mathbf{B}_{v}^{-1}\mathbf{X}$) and then
specializing the result to the internal state for which $\mathbf{l}_{v}=\mathbf{w}$. In contrast to the formulation
presented in Ref. \cite{LatMonCM2015}, the viscous gradient $\mathbf{X}_{v}$
does not remain completely constant when the mapping tensor given in Eq.
(\ref{mapping tensor Ae A}) is calculated (recall that $\mathbf{\mathring{X}%
}_{v}=\mathbf{0}$ implies $\mathbf{\dot{X}}_{v}=\mathbf{wX}_{v}$). As a
result, that mapping tensor does not correspond in general to the partial
gradient of $\mathbf{A}_{e}$ with respect to $\mathbf{A}$ from a
mathematical point of view. This fact will be relevant below. Interestingly, a
clear parallelism between the formulations based on the reversed
multiplicative decomposition and the Sidoroff's one may be established if we
use the symbol $\delta$ (instead of $\partial$) to represent the partial variation of
a non-equilibrated variable constrained by $\mathbf{\mathring{X}}%
_{v}=\mathbf{0}$ (instead of $\mathbf{\dot{X}}_{v}=\mathbf{0}$). Hereafter we
adopt that notation.

Using Eq. (\ref{Aedot Adot}) we obtain the following equivalent material
descriptions of the non-dissipative stress power per unit reference volume
---compare to Eq. (\ref{non-dissipative energy rate})%
\begin{equation}\label{non-dissipative energy rate material}
\mathbf{S}_{neq}^{|e}:\left.  \mathbf{\dot{A}}_{e}\right\vert _{\xvdo}%
=\mathbf{S}_{neq}:\mathbf{\dot{A}}=\left.  \dot{\Psi}_{neq}\right\vert
_{\xvdo}%
\end{equation}
This interpretation gives the existing mapping between the non-equilibrated
Second Piola-Kirchhoff stress tensors $\mathbf{S}_{neq}$ and $\mathbf{S%
}_{neq}^{|e}$
\begin{equation}
\mathbf{S}_{neq}=\mathbf{S}_{neq}^{|e}:\left.  \dfrac{\delta\mathbf{A}_{e}%
}{\delta\mathbf{A}}\right\vert _{\xvdo}=\dfrac{d\Psi_{neq}\left(
\mathbf{A}_{e}\right)  }{d\mathbf{A}_{e}}:\left.  \dfrac{\delta
\mathbf{A}_{e}}{\delta\mathbf{A}}\right\vert _{\xvdo}=\left.  \dfrac
{\delta\Psi_{neq}}{\delta\mathbf{A}%
}\right\vert _{\xvdo} \label{SPK stresses neq}%
\end{equation}
Both stress tensors $\mathbf{S}_{neq}$ and $\mathbf{S}_{neq}^{|e}$ operate in
the reference configuration but they are associated to different deformations,
represented by $\mathbf{A}$ and $\mathbf{A}_{e}$ respectively. Furthermore,
Identity (\ref{SPK stresses neq})$_{3}$ provides the way in which the
non-equilibrated stresses $\mathbf{S}_{neq}$ are obtained from $\Psi_{neq}$ in
this case, i.e. by means of the partial \emph{variation} of $\Psi
_{neq}$ with respect to $\mathbf{A}$\ along the non-dissipative, corotational
path $\mathbf{\mathring{X}}_{v}=\mathbf{0}$.

\section{Finite strain viscoelasticity based on logarithmic strain
measures}

In the preceding section we have obtained all the required tensors needed to
properly formulate a finite fully nonlinear visco-hyperelastic model based on
the reversed decomposition defined in terms of Green-Lagrange
measures. This continuum formulation is valid for anisotropic compressible materials.
However, we are mostly interested in formulating a model for nearly-incompressible materials using logarithmic
strains because of both their special properties \cite{LatMonIJSS2014} and the
possibility of using spline-based stored energy functions \cite{SusBat2009,LatMonCAS2013,LatMonCM2014}.

In order to achieve our objective, it is convenient to decompose first the total deformation gradient using the Flory's decomposition
\begin{equation}
\mathbf{X}=(J^{1/3}\mathbf{I})\mathbf{X}^d
\label{Flory}
\end{equation}
where $\det(\mathbf{X}^d)=1$, and, subsequently, decompose the distortional part of the deformation gradient
by means of the reversed decomposition
\begin{equation}\label{Reversed decomposition deviatoric}
\mathbf{X}^d=\mathbf{X}_v^d\mathbf{X}_e^d\equiv\mathbf{X}_v\mathbf{X}_e
\end{equation}
That way, the isochoric nature of the non-equilibrium part is exactly preserved by construction.

As it is usual when modelling the mechanical behavior of
(nearly-)incompressible materials with application in finite element
procedures, Eq. (\ref{strain energy function A}) is divided into
uncoupled deviatoric and volumetric parts. In terms of the material logarithmic strain
measures associated the preceding multiplicative decompositions, it reads%
\begin{equation}
\Psi=\mathcal{W}+\mathcal{U}=\mathcal{W}_{eq}(\mathbf{E}^{d})+\mathcal{W}%
_{neq}(\mathbf{E}_{e}^{d})+\mathcal{U}_{eq}\left(  J\right)
\label{strain enegy function}%
\end{equation}
where both $\mathbf{E}$ and $\mathbf{E}_{e}^d\equiv\mathbf{E}_{e}$ are defined in the reference
configuration as%
\begin{equation}
\mathbf{E}=\tfrac{1}{2}\ln\left(  \mathbf{C}\right)  =\tfrac{1}{2}\ln\left(
\mathbf{X}^{T}\mathbf{X}\right)  \text{ and }\mathbf{E}_{e}=\tfrac{1}{2}%
\ln\left(  \mathbf{C}_{e}\right)  =\tfrac{1}{2}\ln\left(  \mathbf{X}_{e}%
^{T}\mathbf{X}_{e}\right)
\end{equation}
with
\begin{equation}
\mathbf{E}^{d}=\mathbf{E}-\tfrac{1}{3}tr\left(  \mathbf{E}\right)  \text{
\ with \ }tr\left(  \mathbf{E}\right)  :=\ln J:=\ln\left(  \det\left(
\mathbf{X}\right)  \right)
\end{equation}
In Eq. (\ref{strain enegy function}), $\mathcal{W}=\mathcal{W}_{eq}+\mathcal{W}_{neq}$ (both $\mathcal{W}%
_{eq}$ and$\ \mathcal{W}_{neq}$ to be determined from experimental data)
depends on deviatoric, true, behaviors only and $\mathcal{U}=\mathcal{U}_{eq}$
will be used to introduce the required volumetric penalty constraint to the deformation
($J\approx1$) in the numerical calculations.

In the next sections we derive the expressions of the second Piola-Kirchhoff
stress tensor $\,^{t+\Delta t}\mathbf{S}$ and the corresponding tangent moduli
$\,^{t+\Delta t}\mathbb{C}$ when the multiplicative decomposition $\,_{0}%
^{t}\mathbf{X}=\,_{0}^{t}J^{1/3}\,_{0}^{t}\mathbf{X}_{v}\,_{0}^{t}\mathbf{X}_{e}$ is known at
$t$ and only the deformation gradient $\,_{\hspace{3.2ex}0}^{t+\Delta
t}\mathbf{X}$ is known at $t+\Delta t$ ---for the incremental formulation we use the notation
given in Ref. \cite{Bathe}. We first address how to compute the
non-equilibrated contribution and then we address the simpler equilibrated
one. Finally, the total stresses and tangent moduli are obtained through
$\,^{t+\Delta t}\mathbf{S}=\,^{t+\Delta t}\mathbf{S}_{eq}+\,^{t+\Delta
t}\mathbf{S}_{neq}$ and $\,^{t+\Delta t}\mathbb{C}=\,^{t+\Delta t}%
\mathbb{C}_{eq}+\,^{t+\Delta t}\mathbb{C}_{neq}$.

\section{Non-equilibrated contribution}

\subsection{Constitutive equation for the viscous
flow\label{Section - Constitutive equation for the viscous flow}}

In order to enforce the physical restriction given in Eq.
(\ref{dissipation inequality spatial})$_{2}$ in the logarithmic strain space,
we rewrite it attending to the purely kinematic power-conjugacy equivalence between the
stress power given by the elastic
deformation rate tensor $\mathbf{d}_{e}$ and the stress power given by the material rate of the elastic
logarithmic strains $\mathbf{\dot{E}}_{e}$
\begin{equation}
-\left.  \mathcal{\dot{W}}_{neq}\right\vert _{\ddo}=-\mathbf{\tau}%
_{neq}^{|e}:\left.  \mathbf{d}_{e}\right\vert _{\ddo}=-\mathbf{T}%
_{neq}^{|e}:\left.  \mathbf{\dot{E}}_{e}\right\vert _{\edo}=-\left.  \mathcal{\dot
{W}}_{neq}\right\vert _{\edo}\geq0 \label{dissipation inequality material}%
\end{equation}
where we define the non-equilibrated purely deviatoric generalized Kirchhoff stresses as%
\begin{equation}
\mathbf{T}_{neq}^{|e}:=\frac{d\mathcal{W}_{neq}}{d
\mathbf{E}_{e}}=\frac{d\mathcal{W}_{neq}}{d\mathbf{E}_{e}^{d}%
}:\frac{d\mathbf{E}_{e}^{d}}{d\mathbf{E}_{e}}=\frac
{d\mathcal{W}_{neq}}{d\mathbf{E}_{e}^{d}}:\mathbb{P}%
^{S} \label{T hat neq}%
\end{equation}
with $\mathbb{P}^{S}=\mathbb{I}^{S}-\frac{1}{3}\mathbf{I}\otimes\mathbf{I}$ being the fourth-order
symmetric deviatoric projection tensor, with components in any given basis
\begin{equation}
\left(  \mathbb{P}^{S}\right)  _{ijkl}=\frac{1}{2}\left(  \delta
_{ik}\delta_{jl}+\delta_{il}\delta_{jk}\right)  -\frac{1}{3}\delta_{ij}%
\delta_{kl}%
\end{equation}
The stress tensor $\mathbf{T}_{neq}^{|e}$ relates to the Kirchhoff stress tensor $\mathbf{\tau}_{neq}^{|e}$ through%
\begin{equation}
\mathbf{T}_{neq}^{|e}=\mathbf{\tau}_{neq}^{|e}:\mathbb{M}_{\dot{E}_{e}%
}^{d_{e}}%
\end{equation}
where the mapping tensor $\mathbb{M}_{\dot{E}_{e}}^{d_{e}}$\ (not needed herein)
may be easily obtained in spectral form \cite{LatMonCM2015}. Equation
(\ref{dissipation inequality material}) is automatically satisfied if we
choose the following flow rule%
\begin{equation}
-\left.  \frac{d\mathbf{E}_{e}}{dt}\right\vert _{\edo}=\mathbb{V}%
^{-1}:\mathbf{T}_{neq}^{|e} \label{viscous flow rule}%
\end{equation}
for a given fourth-order positive-definite viscosity tensor $\mathbb{V}^{-1}$,
whereupon%
\begin{equation}\label{possitiveDissipation}
\mathbf{T}_{neq}^{|e}:\mathbb{V}^{-1}:\mathbf{T}_{neq}^{|e}\geq0
\end{equation}
As an important difference with respect to the models based on the Sidoroff
decomposition, note that all the entities present in Eqs.
(\ref{viscous flow rule}) and (\ref{possitiveDissipation}) are defined in the reference configuration.

\subsection{Integration of the evolution equation}

\begin{figure}
[ptb]
\begin{center}
\includegraphics[width=1.\textwidth]%
{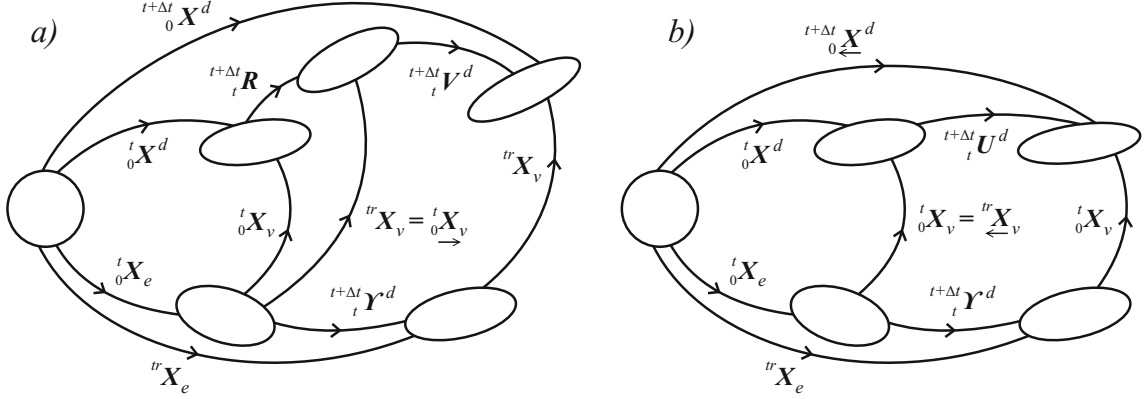}%
\caption{Multiplicative decomposition of the (isochoric) trial state at $t+\Delta t$. Two
equivalent interpretations.}%
\label{figure03.eps}%
\end{center}
\end{figure}

The non-linear viscous flow rule given in Eq. (\ref{viscous flow rule}) can
be integrated by means of a two-step, elastic predictor/viscous corrector
incremental scheme. Within the elastic predictor substep there is no viscous
dissipation, so Eq. (\ref{dissipation inequality spatial})$_{1}$ yields%
\begin{equation}
\,^{tr}\mathbf{l}_{v}^{\circ}=\mathbf{0}\quad\Rightarrow\quad\,^{tr}%
\mathbf{l}_{v}=\,^{t+\Delta t}\mathbf{w}\quad\Rightarrow\quad\,^{tr}%
\mathbf{\dot{X}}_{v}=\,^{t+\Delta t}\mathbf{w}\,^{tr}\mathbf{X}_{v}%
\end{equation}
which may be integrated, employing the usual exponential mapping%
\begin{equation}
\,^{tr}\mathbf{X}_{v}=\exp\left(  \,^{t+\Delta t}\mathbf{w}\Delta t\right)
\,_{0}^{t}\mathbf{X}_{v} \label{Xv trial}%
\end{equation}
The tensor $\exp(\,^{t+\Delta t}\mathbf{w}\Delta t)$ can be identified after
the integration of the equation $\mathbf{\dot{X}}=\mathbf{lX}$, i.e.%
\begin{equation}
_{\hspace{3.2ex}0}^{t+\Delta t}\mathbf{X}=\exp\left(  \,^{t+\Delta
t}\mathbf{l}\Delta t\right)  \,_{0}^{t}\mathbf{X}\approx\exp\left(
\,^{t+\Delta t}\mathbf{d}\Delta t\right)  \exp\left(  \,^{t+\Delta
t}\mathbf{w}\Delta t\right)  \,_{0}^{t}\mathbf{X}%
\end{equation}
and then comparing this approximation to the incremental multiplicative
decomposition%
\begin{equation}
\,_{\hspace{3.2ex}0}^{t+\Delta t}\mathbf{X}=\,_{\hspace{3.2ex}t}^{t+\Delta
t}\mathbf{X}\,_{0}^{t}\mathbf{X}=\,_{\hspace{3.2ex}t}^{t+\Delta t}%
\mathbf{V}\,_{\hspace{3.2ex}t}^{t+\Delta t}\mathbf{R}\,_{0}^{t}\mathbf{X}%
\end{equation}
where $\,_{\hspace{3.2ex}t}^{t+\Delta t}\mathbf{V}$ and $\,_{\hspace{3.2ex}%
t}^{t+\Delta t}\mathbf{R}$ are the stretch and rotation tensors from the left
polar decomposition of the incremental deformation gradient $\,_{\hspace
{3.2ex}t}^{t+\Delta t}\mathbf{X}$ relating the configurations at $t$ and
$t+\Delta t$. Note that, in general, $\,_{\hspace{3.2ex}0}^{t+\Delta
t}\mathbf{V}\neq\,_{\hspace{3.2ex}t}^{t+\Delta t}\mathbf{V}\,_{0}%
^{t}\mathbf{V}$ and $\,_{\hspace{3.2ex}0}^{t+\Delta t}\mathbf{R}%
\neq\,_{\hspace{3.2ex}t}^{t+\Delta t}\mathbf{R}\,_{0}^{t}\mathbf{R}$, see discussion in Ref. \cite{MontansBenitezCaminero}. However, $\,_{\hspace{3.2ex}0}^{t+\Delta t}J%
=\,_{\hspace{3.2ex}t}^{t+\Delta t}J\,_{0}^{t}J$. Hence,
we can approximate the incremental distortional deformation by means of%
\begin{equation}
\exp\left(  \,^{t+\Delta t}\mathbf{d}^d\Delta t\right)  =\,_{\hspace{3.2ex}%
t}^{t+\Delta t}\mathbf{V}^d\text{\qquad and\qquad}\exp\left(  \,^{t+\Delta
t}\mathbf{w}\Delta t\right)  =\,_{\hspace{3.2ex}t}^{t+\Delta t}\mathbf{R}
\label{V R t t+dt}%
\end{equation}
Equations (\ref{Xv trial}) and (\ref{V R t t+dt})$_{2}$ provide the definition
of the isochoric trial state at time $t+\Delta t$ as ---the right arrow decoration means
\textquotedblleft rotated by $\,_{\hspace{3.2ex}t}^{t+\Delta t}\mathbf{R}%
$\textquotedblright%
\begin{align}
\,^{tr}\mathbf{X}_{v}  &  =\,_{\hspace{3.2ex}t}^{t+\Delta t}\mathbf{R}%
\,_{0}^{t}\mathbf{X}_{v}=:\,_{0}^{t}\underrightarrow{\mathbf{X}_{v}}\\
\,^{tr}\mathbf{X}_{e}  &  =\,^{tr}\mathbf{X}_{v}^{-1}\,_{\hspace{3.2ex}%
0}^{t+\Delta t}\mathbf{X}^d=(\,_{0}^{t}\underrightarrow{\mathbf{X}_{v}^{-1}%
}\,_{\hspace{3.2ex}t}^{t+\Delta t}\mathbf{V}^d\,_{0}^{t}\underrightarrow
{\mathbf{X}_{v}})\,_{0}^{t}\mathbf{X}_{e}=\,_{\hspace{3.2ex}t}^{t+\Delta
t}\mathbf{\Upsilon}^d\,_{0}^{t}\mathbf{X}_{e} \label{Xetr 1}%
\end{align}
where all the quantities needed for the calculation of $\,^{tr}\mathbf{X}_{v}$ and
$\,^{tr}\mathbf{X}_{e}$\ are known. Note that trial states are defined in base
of Eq. (\ref{Xv trial}) and have different form from usual set-ups based on
the Sidoroff decomposition \cite{Sidoroff,ReeseGovindjee,LatMonCM2015} or the Lee decomposition in plasticity
\cite{Lee,EterovicBathe,CamineroMontansBathe}. Hence, in this case, we may
interpret that the increment of isochoric deformation $\,_{\hspace{3.2ex}%
t}^{t+\Delta t}\mathbf{V}^d$ is completely applied to the elastic deformation
gradient $\,_{0}^{t}\mathbf{\mathbf{X}}_{e}$ within the trial substep by means
of the pull-back of $\,_{\hspace{3.2ex}t}^{t+\Delta t}\mathbf{V}^d$ to the
intermediate configuration through, see Figure \ref{figure03.eps}.a%
\begin{equation}
\,_{\hspace{3.2ex}t}^{t+\Delta t}\mathbf{\Upsilon}^d:=\,_{0}^{t}\underrightarrow
{\mathbf{X}_{v}^{-1}}\,_{\hspace{3.2ex}t}^{t+\Delta t}\mathbf{V}^d\,_{0}%
^{t}\underrightarrow{\mathbf{X}_{v}}
\label{Vpulled}%
\end{equation}

A more intuitive interpretation is obtained if we previously rotate the actual
configuration at $t+\Delta t$ with $\,_{\hspace{3.2ex}t}^{t+\Delta
t}\mathbf{R}^{T}$ (i.e. if we remove the rotation $\,_{\hspace{3.2ex}%
t}^{t+\Delta t}\mathbf{R}$ from the two-point total and viscous deformation gradient tensors) as shown in Figure \ref{figure03.eps}.b. In that case we
have ---the left arrow decoration means \textquotedblleft rotated by
$\,_{\hspace{3.2ex}t}^{t+\Delta t}\mathbf{R}^{T}$\textquotedblright%
\begin{align}
\,_{\hspace{3.2ex}0}^{t+\Delta t}\underleftarrow{\mathbf{X}^d}  &
:=\,_{\hspace{3.2ex}t}^{t+\Delta t}\mathbf{R}^{T}\,_{\hspace{3.2ex}%
0}^{t+\Delta t}\mathbf{X}^d=\,_{\hspace{3.2ex}t}^{t+\Delta t}\mathbf{U}^d%
\,_{0}^{t}\mathbf{X}^d\\
\,^{tr}\underleftarrow{\mathbf{X}_{v}}  &  :=\,_{\hspace{3.2ex}t}^{t+\Delta
t}\mathbf{R}^{T}\,^{tr}\mathbf{X}_{v}=\,_{0}^{t}\mathbf{X}_{v}\\
\,^{tr}\mathbf{X}_{e}  &  =\,^{tr}\underleftarrow{\mathbf{X}_{v}^{-1}%
}\,_{\hspace{3.2ex}0}^{t+\Delta t}\underleftarrow{\mathbf{X}^d}=(\,_{0}%
^{t}\mathbf{X}_{v}^{-1}\,_{\hspace{3.2ex}t}^{t+\Delta t}\mathbf{U}^d\,_{0}%
^{t}\mathbf{X}_{v})\,_{0}^{t}\mathbf{X}_{e}=\,_{\hspace{3.2ex}t}^{t+\Delta
t}\mathbf{\Upsilon}^d\,_{0}^{t}\mathbf{X}_{e} \label{Xetr 2}%
\end{align}
where$\ \,_{\hspace{3.2ex}t}^{t+\Delta t}\mathbf{U}^d\approx\exp(\,^{t+\Delta
t}\underleftarrow{\mathbf{d}^d}\Delta t)$ is the distortional stretch tensor from the right
polar decomposition of$\ \,_{\hspace{3.2ex}t}^{t+\Delta t}\mathbf{X}^d$. Thus we
observe that, equivalently, the increment of isochoric deformation $\,_{\hspace
{3.2ex}t}^{t+\Delta t}\mathbf{U}^d$ is completely applied to the elastic
deformation gradient $\,_{0}^{t}\mathbf{\mathbf{X}}_{e}$ within the trial
substep by means of the pull-back of$\ \,_{\hspace{3.2ex}t}^{t+\Delta
t}\mathbf{U}^d$ to the intermediate configuration through ---c.f. Eq.(\ref{Vpulled})%
\begin{equation}
\,_{\hspace{3.2ex}t}^{t+\Delta t}\mathbf{\Upsilon}^d=\,_{0}^{t}\mathbf{X}%
_{v}^{-1}\,_{\hspace{3.2ex}t}^{t+\Delta t}\mathbf{U}^d\,_{0}^{t}\mathbf{X}_{v}%
\end{equation}
In any case, the trial logarithmic strain tensor is%
\begin{equation}
\,^{tr}\mathbf{E}_{e}=\dfrac{1}{2}\ln(\,^{tr}\mathbf{C}_{e})=\dfrac{1}{2}%
\ln(\,^{tr}\mathbf{X}_{e}^{T}\,^{tr}\mathbf{X}_{e})
\label{trial elastic logarithmic strains}%
\end{equation}
with $\,^{tr}\mathbf{X}_{e}=\,_{0}^{t}\underrightarrow{\mathbf{X}_{v}^{-1}%
}\,_{\hspace{3.2ex}0}^{t+\Delta t}\mathbf{X}^d$ or $\,^{tr}\mathbf{X}_{e}%
=\,_{0}^{t}\mathbf{X}_{v}^{-1}\,_{\hspace{3.2ex}0}^{t+\Delta t}\underleftarrow
{\mathbf{X}^d}$.

Trial states associated to the models based on either the Sidoroff's
decomposition or the reversed one are different in general. The respective
trial states, and hence the respective integration algorithms, are coincident
for the very special cases of axial loadings in isotropic materials or in
orthotropic materials along the preferred material directions.

Subsequently, during the viscous corrector substep the total deformation rate
$\mathbf{d} = 0 $. We exactly proceed as in Ref. \cite{LatMonCM2015}%
\begin{equation}
\mathbf{\dot{E}}=\mathbf{0\quad\Rightarrow\quad\mathbf{\dot{E}}}_{e}=\left.
\mathbf{\mathbf{\dot{E}}}_{e}\right\vert _{\edo}%
\end{equation}
i.e. using a backward-Euler integration in Eq. (\ref{viscous flow rule})%
\begin{equation}
\,_{\hspace{3.2ex}0}^{t+\Delta t}\mathbf{E}_{e}-\,^{tr}\mathbf{E}_{e}%
\approx-\Delta t\left(  \mathbb{V}^{-1}:\mathbf{T}_{neq}^{|e}\right)
_{t+\Delta t}%
\end{equation}
which provides a non-linear viscous correction for $\,_{\hspace{3.2ex}%
0}^{t+\Delta t}\mathbf{E}_{e}$ in terms of $\,^{tr}\mathbf{E}_{e}$ through%
\begin{equation}
\,_{\hspace{3.2ex}0}^{t+\Delta t}\mathbf{E}_{e}+\Delta t\left(  \mathbb{V}%
^{-1}:\dfrac{d\mathcal{W}_{neq}}{d\mathbf{E}_{e}}\right)
_{t+\Delta t}=\,^{tr}\mathbf{E}_{e} \label{viscous correction}%
\end{equation}
The same non-linear evolution equation is obtained for the model based on the
Sidoroff's decomposition that we derived in Ref. \cite{LatMonCM2015}. In both
frameworks, once $\mathbb{V}^{-1}$ and $\mathcal{W}_{neq}$ are known, we can
compute $\,_{\hspace{3.2ex}0}^{t+\Delta t}\mathbf{E}_{e}$ for a given time
step $\Delta t$ performing local iterations at the integration point level and
then proceed to obtain the deviatoric non-equilibrated stresses and tangent
moduli at $t+\Delta t$. However, two differences of distinct nature have to be
emphasized regarding the update of Eq. (\ref{viscous correction}) associated
to either the Sidoroff or to the reverse decompositions. On the one hand, we
have just seen that the numerical calculation of the trial state is different
for both decompositions and that the respective trial states are only
coincident in very specific cases. On the other hand, upon the acceptance of
the reversed kinematic decomposition, we have seen that Eq.
(\ref{viscous correction}) is completely defined in the reference
configuration. Hence, no further hypothesis regarding the evolution of the
preferred directions are required in this case because the non-equilibrated
strain energy function $\mathcal{W}_{neq}$ is defined in the reference
configuration and the corresponding Lagrangian non-equilibrated stresses are
properly derived in that configuration, whatever the material symmetries are.
The tensor $\mathbb{V}^{-1}$ may also be defined with the same material
symmetries of $\mathcal{W}_{neq}$.

For practical purposes but without loss of generality of the present formulation, we will
assume herein that $\mathbb{V}^{-1}$ is a purely deviatoric orthotropic tensor
given in terms of six scalar viscosity parameters $\eta_{ij}^{d}=\eta_{ji}^{d}$ through%
\begin{equation}
\mathbb{V}^{-1}=\mathbb{P}^{S}:\underset{%
\begin{array}
[c]{c}%
\mathbb{\bar{V}}^{-1}%
\end{array}
}{\ \underbrace{\left(
{\displaystyle\sum\limits_{i=1}^{3}}
{\displaystyle\sum\limits_{j=1}^{3}}
\frac{1}{2\eta_{ij}^{d}}\mathbf{L}_{ij}^{S}\otimes\mathbf{L}_{ij}^{S}\right)
}\ }:\mathbb{P}^{S}%
\label{viscosity tensor}%
\end{equation}
where $\mathbf{L}_{ij}^{S}=1/2(\mathbf{a}_{i}\otimes\mathbf{a}_{j}%
+\mathbf{a}_{j}\otimes\mathbf{e}_{i})$ stand for the structural tensors
associated to the material preferred basis $X_{pr}=\{\mathbf{a}_{1},\mathbf{a}%
_{2},\mathbf{a}_{3}\}$.
The evolution equation in rate form Eq. (\ref{viscous flow rule}) and its
solution in terms of incremental elastic strains Eq. (\ref{viscous correction})
become purely deviatoric. It is apparent that, in
general, $\,_{\hspace{3.2ex}0}^{t+\Delta t}\mathbf{E}_{e}$ and $\,^{tr}%
\mathbf{E}_{e}$ in Eq. (\ref{viscous correction}) will not have the same
Lagrangian principal basis. In Section
\ref{Section - Determination of the relaxation time(s) of the orthotropic model}
we show how to obtain the values of the material parameters $\eta_{ij}^{d}$
from experimental testing. Once the viscosity parameters $\eta_{ij}^{d}$
are known, the non-linear Equations (\ref{viscous flow rule}) and
(\ref{viscous correction}) are to be used. In those equations we will further
assume that the viscosity parameters are deformation independent.

The value of the material parameters $\eta_{ij}^{d}$ in Eq.
(\ref{viscosity tensor}) may be related to a set of six independent
relaxation times $\tau_{ij}=\tau_{ji}$ for the orthotropic case. Note that we use the same symbol for
the relaxation times as for the Kirchhoff stresses but by context confusion
is hardly possible.

In order to obtain the existing relations between $\eta_{ij}^{d}$ and
$\tau_{ij}$\ we must linearize the response of the non-equilibrated
orthotropic strain energy function $\mathcal{W}_{neq}$ in the flow rule of Eq.
(\ref{viscous flow rule}) to obtain%
\begin{equation}
-\left.  \frac{d\mathbf{E}_{e}}{dt}\right\vert _{\edo}=\mathbb{P}%
^{S}:\underset{%
\begin{array}
[c]{c}%
\mathbb{\bar{T}}_{lin}^{-1}%
\end{array}
}{\ \underbrace{\left(  \mathbb{\bar{V}}^{-1}:\mathbb{P}^{S}:\left.
\dfrac{d^{2}\mathcal{W}_{neq}}{d\mathbf{E}_{e}^{d}%
d\mathbf{E}_{e}^{d}}\right\vert _{lin}\right)  }\ }:\mathbb{P}%
^{S}:\mathbf{E}_{e}=\left.  \mathbb{T}_{d}^{-1}\right\vert _{lin}%
:\mathbf{E}_{e} \label{viscous flow rule linearized}%
\end{equation}
where%
\begin{equation}
\left.  \mathcal{W}_{neq}(\mathbf{E}_{e}^{d})\right\vert _{lin}=%
{\displaystyle\sum\limits_{i=1}^{3}}
{\displaystyle\sum\limits_{j=1}^{3}}
\mu_{ij}^{neq}(\mathbf{a}_{i}\cdot\mathbf{E}_{e}^{d}\mathbf{a}_{j})^{2}=%
{\displaystyle\sum\limits_{i=1}^{3}}
{\displaystyle\sum\limits_{j=1}^{3}}
\mu_{ij}^{neq}(E_{eij}^{d})^{2} \label{Wneq lin}%
\end{equation}
is expressed in terms of the orthotropic reference shear moduli $\mu
_{ij}^{neq}$ and the components of $\mathbf{E}_{e}^{d}$ in the material
orthotropy basis $X_{pr}=\left\{  \mathbf{a}_{1},\mathbf{a}_{2},\mathbf{a}%
_{3}\right\}  $. The subscript $lin$ implies a linearized constitutive law
(usually at the origin), i.e. quadratic strain energy with constant
coefficients. The linearized fourth-order deviatoric \emph{relaxation}\ tensor
$\left.  \mathbb{T}_{d}^{-1}\right\vert _{lin}$ present in Eq.
(\ref{viscous flow rule linearized}) is given in terms of the tensor
$\mathbb{\bar{T}}_{lin}^{-1}$, whose matrix (Voigt) representation in the
preferred axes $X_{pr}$ is%
\begin{equation}
\left[  \mathbb{\bar{T}}_{lin}^{-1}\right]  _{X_{pr}}=\left[
\begin{array}
[c]{cccccc}%
\dfrac{2}{3}\dfrac{1}{\tau_{11}} & -\dfrac{1}{3}\dfrac{\rho_{21}}{\tau_{22}} &
-\dfrac{1}{3}\dfrac{\rho_{31}}{\tau_{33}} & 0 & 0 & 0 \medskip\\
-\dfrac{1}{3}\dfrac{\rho_{12}}{\tau_{11}} & \dfrac{2}{3}\dfrac{1}{\tau_{22}} &
-\dfrac{1}{3}\dfrac{\rho_{32}}{\tau_{33}} & 0 & 0 & 0 \medskip\\
-\dfrac{1}{3}\dfrac{\rho_{13}}{\tau_{11}} & -\dfrac{1}{3}\dfrac{\rho_{23}%
}{\tau_{22}} & \dfrac{2}{3}\dfrac{1}{\tau_{33}} & 0 & 0 & 0\\
0 & 0 & 0 & \dfrac{1}{\tau_{12}} & 0 & 0\\
0 & 0 & 0 & 0 & \dfrac{1}{\tau_{23}} & 0\\
0 & 0 & 0 & 0 & 0 & \dfrac{1}{\tau_{31}}%
\end{array}
\right]  \label{viscosity tensor ortho}%
\end{equation}
The relaxation times $\tau_{ij}$ and the coupling coefficients $\rho
_{ij}$ are given by%
\begin{align}
\tau_{ij}  &  :=\frac{\eta_{ij}^{d}}{\mu_{ij}^{neq}}~,\quad
i,~j=\{1,2,3\}\label{relaxation times 0}\\
\rho_{ij}  &  :=\dfrac{\eta_{ii}^{d}}{\eta_{jj}^{d}}~,\quad i\neq j=\{1,2,3\}
\label{couplings 0}%
\end{align}
The tensor $\mathbb{\bar{T}}_{lin}^{-1}$, as given in Eq.
(\ref{viscosity tensor ortho}), is non-symmetric in general. We remark that,
as a main difference with the isotropic viscosity/orthotropic elasticity
formulation presented in Ref. \cite{LatMonCM2015}, the six relaxation times
given in Eq. (\ref{relaxation times 0}) are completely independent, hence
leading to a more general anisotropic visco-hyperelasticity formulation. We
show the high (enhanced) versatility of the present model in the examples
below. If the viscosity tensor is regarded as isotropic in Eq.
(\ref{viscosity tensor}), then we can write $\mathbb{V}^{-1}=1/(2\eta^{d}%
)\mathbb{I}^{S}$ and the update formula Eq. (\ref{viscous correction}) adopts
the same form as in Ref. \cite{LatMonCM2015}, even though it is formulated
herein in the reference configuration. Equation (\ref{viscosity tensor ortho})
reduces in this case to $\mathbb{\bar{T}}_{lin}^{-1}=\mathbb{P}^{S}:\mathbb{T}%
_{lin}^{-1}$, with $\mathbb{T}_{lin}^{-1}$ resulting in a tensor with diagonal matrix representation, which was for
simplicity the shape chosen in Reference \cite{LatMonCM2015} for isotropic
viscous behavior.

Finally, once a converged solution $\,_{\hspace{3.2ex}0}^{t+\Delta
t}\mathbf{E}_{e}$ has been obtained from Eq. (\ref{viscous correction}), the following (internal volume-preserving) update may be performed%
\begin{equation}
\,_{\hspace{3.2ex}0}^{t+\Delta t}\mathbf{X}_{e}=\,^{tr}\mathbf{R}%
_{e}\,_{\hspace{3.2ex}0}^{t+\Delta t}\mathbf{U}_{e}=\,^{tr}\mathbf{R}_{e}%
\exp\left(  \,_{\hspace{3.2ex}0}^{t+\Delta t}\mathbf{E}_{e}\right)
\end{equation}
or%
\begin{equation}
\,_{\hspace{3.2ex}0}^{t+\Delta t}\mathbf{X}_{v}=\,_{\hspace{3.2ex}0}^{t+\Delta
t}\mathbf{X}^d\,_{\hspace{3.2ex}0}^{t+\Delta t}\mathbf{X}_{e}^{-1}%
=\,^{tr}\mathbf{X}_{v}\left(  \,^{tr}\mathbf{V}_{e}\,_{\hspace{3.2ex}0}^{t+\Delta
t}\mathbf{V}_{e}^{-1}\right)%
\end{equation}

\subsection{Local Newton iterations for the non-equilibrated
part\label{Section - Local Newton iterations for the non-equilibrated part}}

Once the trial elastic logarithmic strains $\,^{tr}\mathbf{E}_{e}$ have been
obtained using Eq. (\ref{trial elastic logarithmic strains}), we proceed to
solve Eq. (\ref{viscous correction}) in residual form
for the most general case when hyperelasticity is non-linear in logarithmic
strains. We can proceed as in Ref. \cite{LatMonCM2015} but considering
the residual equation%
\begin{equation}
\mathbf{R}^{(k)}_{\mathbf{ _E}}=\mathbf{E}_{e}^{(k)}+\Delta t\left(  \mathbb{V}%
^{-1}:\left.  \dfrac{d\mathcal{W}_{neq}}{d\mathbf{E}_{e}%
}\right\vert _{\left(  k\right)  }\right)  -\,^{tr}\mathbf{E}_{e}
\label{residual}%
\end{equation}
and employing the \emph{non-symmetric} gradient
\begin{equation}
\frac{d\mathbf{R_E}}{d\mathbf{E}_{e}}=\mathbb{I}^{S}+\Delta
t\left(  \mathbb{V}^{-1}:\dfrac{d^{2}\mathcal{W}_{neq}}%
{d\mathbf{E}_{e}d\mathbf{E}_{e}}\right)  =\mathbb{I}%
^{S}+\mathbb{P}^{S}:\Delta t\left(  \mathbb{\bar{V}}^{-1}:\mathbb{P}%
^{S}:\dfrac{d^{2}\mathcal{W}_{neq}}{d\mathbf{E}_{e}^{d}%
d\mathbf{E}_{e}^{d}}\right)  :\mathbb{P}^{S} \label{dR/dEe}%
\end{equation}
The resulting iterative procedure for $\mathbf{E}_{e}$ is volume-preserving due to the fact that the term between
parenthesis in Eq. (\ref{residual}) is purely deviatoric.

\subsection{Non-equilibrated contribution to $\mathbf{S}$ and $\mathbb{C}%
$%
\label{Section - Deviatoric non-equilibrated contribution to stress and tangent}%
}

Once the elastic strains $\mathbf{E}_{e}$ are known at $t+\Delta t$ we can
proceed to compute the deviatoric non-equilibrated contribution to the stress
and global tangent tensors. As we did in Ref. \cite{LatMonCM2015}, it is
convenient to take derivatives with respect to trial quantities in order to obtain
the non-equilibrated stresses and tangent moduli consistent with the predictor/corrector integration algorithm
employed and then perform the corresponding mappings.

First of all, the consideration of the Flory's decomposition of Eq. (\ref{Flory}) in Eq. (\ref{SPK stresses neq})$_3$ gives
\begin{equation}\label{S_neq_dev}
\mathbf{S}_{neq}=\left.  \dfrac{\delta\mathcal{W}_{neq}}{\delta
\mathbf{A}}\right\vert _{\xvdo}=\left.  \dfrac{\delta\mathcal{W}_{neq}}{\delta
\mathbf{A}^{d}}\right\vert _{\xvdo}:\dfrac{d%
\mathbf{A}^{d}}{d\mathbf{A}}:=\mathbf{S}_{neq}^{|d}:\dfrac{d%
\mathbf{A}^{d}}{d\mathbf{A}}
\end{equation}
where $\mathbf{S}_{neq}^{|d}$ is a modified second Piola--Kirchhoff stress tensor
defined in the reference configuration and $d\mathbf{A}^{d}/d\mathbf{A}$ represents the fourth-order deviatoric
projection tensor in the space of quadratic strains, see below.

The trial state is defined by $\,^{tr}\mathbf{l}_{v}^{\circ}=\mathbf{0}$. Then
the distortional counterpart of Eqs. (\ref{deformation rates}), (\ref{deformation rates 2}) and
(\ref{mapping tensor d de}) particularized to the trial state read ---note
that subscripts of the type $\,^{tr}\mathbf{l}_{v}^{\circ}=\mathbf{0}$ or $\,^{tr}\mathring{\mathbf{X}}_v=\mathbf{0}$ would be
redundant for the trial state, hence they are not indicated in the corresponding mapping tensors
\begin{equation}
\,^{tr}\mathbf{d}_{e}=sym\left(  \,^{tr}\mathbf{X}_{v}^{-1}\mathbf{d}^d%
\,^{tr}\mathbf{X}_{v}\right)  =\,^{tr}\mathbf{X}_{v}^{-1}\overset{s}{\odot
}\,^{tr}\mathbf{X}_{v}^{T}:\mathbf{d}^d=\mathbb{M}_{{d}^d}^{\,^{tr}d_{e}}:\mathbf{d}^d
\label{detr d}%
\end{equation}
As done in Section \ref{Section - Material description}, the
Lagrangian description of Eq. (\ref{detr d}) is obtained as%
\begin{equation}
\,^{tr}\mathbf{\dot{A}}_{e}=\,^{tr}\mathbf{C}_{e}\mathbf{C}^{d-1}\overset
{s}{\odot}\mathbf{I}:\mathbf{\dot{A}}^d=\dfrac{\delta\,^{tr}\mathbf{A}_{e}%
}{\delta\mathbf{A}^d}:\mathbf{\dot{A}}^d \label{Aetr dot A dot}%
\end{equation}
which gives the mapping associated to the change of the independent variable
$\mathbf{A}^d$ by the independent variable $\,^{tr}\mathbf{A}_{e}$.
We define now the non-equilibrated
second Piola-Kirchhoff stress tensor $\mathbf{S}_{neq}^{|tr}$ associated to the trial state, which operates
in the reference configuration as well, such that ---recall also Eq. (\ref{non-dissipative energy rate material})
\begin{equation}
\mathbf{S}_{neq}^{|tr}:\,^{tr}\mathbf{\dot{A}}_{e}=\mathbf{S}_{neq}^{|d}%
:\mathbf{\dot{A}}^d=\mathbf{S}_{neq}:\mathbf{\dot{A}}=\left.  \mathcal{\dot{W}}_{neq}\right\vert _{\xvdo}%
\end{equation}
which gives the following relation between the non-equilibrated stress tensors
$\mathbf{S}_{neq}^{|d}$ and $\mathbf{S}_{neq}^{|tr}$
\begin{equation}
\mathbf{S}_{neq}^{|d}=\mathbf{S}_{neq}^{|tr}:\,^{tr}\mathbf{C}_{e}\mathbf{C}%
^{d-1}\overset{s}{\odot}\mathbf{I}=\mathbf{S}_{neq}^{|tr}:\dfrac
{\delta\,^{tr}\mathbf{A}_{e}}{\delta\mathbf{A}^d} \label{S neq 0}%
\end{equation}

One important difference between this algorithmic formulation and the one
presented in Ref. \cite{LatMonCM2015} is that in this case the trial
intermediate configuration does not remain constant during the finite-element
global iterations at time $t+\Delta t$ because this configuration is given by
the trial elastic internal gradient, see Eqs. (\ref{Xetr 1}) or (\ref{Xetr 2}%
). Hence, the fourth-order mapping tensor $\delta\,^{tr}\mathbf{A}_{e}%
/\delta\mathbf{A}^d$ present in Eq. (\ref{S neq 0}) has also to be
differentiated in order to obtain the existing relation between the consistent
tangent moduli $\mathbb{C}_{neq}^{|d}=d\mathbf{S}_{neq}^{|d}/d\mathbf{A}^d$
and $\mathbb{C}_{neq}^{|tr}=d\mathbf{S}_{neq}^{|tr}/d
\,^{tr}\mathbf{A}_{e}$, which are to be obtained taking the total derivatives
of $\mathbf{S}_{neq}^{|d}$ and $\mathbf{S}_{neq}^{|tr}$ with respect to
$\mathbf{A}^d$ and $\,^{tr}\mathbf{A}_{e}$ respectively, see discussion in
Ref. \cite{LatMonCM2015}. In this case we have ---note that $\delta
\,^{tr}\mathbf{A}_{e}/\delta\mathbf{A}^d$ has only minor symmetries and that the
second addend in the right-hand side of the following equation vanishes in the model based on the
Sidoroff decomposition%
\begin{equation}
\dfrac{d\mathbf{S}_{neq}^{|d}}{dt}=\dfrac{d\mathbf{S}_{neq}^{|tr}}{dt}%
:\dfrac{\delta\,^{tr}\mathbf{A}_{e}}{\delta\mathbf{A}^d}+\mathbf{S%
}_{neq}^{|tr}:\dfrac{d}{dt}\left(  \dfrac{\delta\,^{tr}\mathbf{A}_{e}}%
{\delta\mathbf{A}^d}\right)
\end{equation}
Expressing the preceding equation in terms of $\mathbf{\dot{A}}^d$ and
$\,^{tr}\mathbf{\dot{A}}_{e}$, using Eq. (\ref{Aetr dot A dot}) and
identifying terms%
\begin{equation}
\dfrac{d\mathbf{S}_{neq}^{|d}}{d\mathbf{A}^d}=\left(  \dfrac
{\delta\,^{tr}\mathbf{A}_{e}}{\delta\mathbf{A}^d}\right)  ^{T}:\dfrac
{d\mathbf{S}_{neq}^{|tr}}{d\,^{tr}\mathbf{A}_{e}}%
:\dfrac{\delta\,^{tr}\mathbf{A}_{e}}{\delta\mathbf{A}^d}+\mathbf{S%
}_{neq}^{|tr}:\dfrac{d}{d\mathbf{A}^d}\left(  \dfrac{\delta
\,^{tr}\mathbf{A}_{e}}{\delta\mathbf{A}^d}\right)  \label{C neq 00}%
\end{equation}
which, after some lengthy algebra, results in%
\begin{align}
\mathbb{C}_{neq}^{|d}&=\left(  \dfrac{\delta\,^{tr}\mathbf{A}_{e}}{\delta\mathbf{A}^d%
}\right)  ^{T}:\mathbb{C}_{neq}^{|tr}:\dfrac{\delta\,^{tr}\mathbf{A}_{e}%
}{\delta\mathbf{A}^d}\nonumber\\
&+\mathbf{C}^{d-1}\,^{tr}\mathbf{C}_{e}\mathbf{C}%
^{d-1}\overset{s}{\odot}\mathbf{S}_{neq}^{|tr}-\mathbf{C}^{d-1}\overset
{s}{\odot}\mathbf{S}_{neq}^{|tr}\,^{tr}\mathbf{C}_{e}\mathbf{C}^{d-1}
\label{C neq 0}%
\end{align}
The first and second addends in the right-hand side of Eq. (\ref{C neq 0})
have major symmetries, the former due to the major symmetry of $\mathbb{
C}_{neq}^{|tr}$ (see below) and the latter due to the symmetry of the second order
tensors $\mathbf{C}^{d-1}\,^{tr}\mathbf{C}%
_{e}\mathbf{C}^{d-1}$ and $\mathbf{S}_{neq}^{|tr}$. However, the third addend in the right-hand side of Eq.
(\ref{C neq 0}) lacks major symmetry, in general. As a result, the tangent
moduli tensor $\mathbb{C}_{neq}^{|d}$ may be slightly non-symmetric, as we show in
the examples. The lack of major symmetry in general situations emerges
from the fact that the fourth-order tensor $\delta\,^{tr}\mathbf{A}_{e}%
/\delta\mathbf{A}^d$ present in Eq. (\ref{C neq 00}) does not exactly correspond
to the gradient of $\,^{tr}\mathbf{A}_{e}$ with respect to $\mathbf{A}^d$, as we
have equivalently explained in Section \ref{Section - Material description}
using the strain tensors $\mathbf{A}_{e}$ and $\mathbf{A}$ and the constraint
$\mathbf{\mathring{X}}_{v}=\mathbf{0}$, recall also Eq. (\ref{mapping tensor d de}). In other words, note that an explicit
expression that gives $\,^{tr}\mathbf{A}_{e}$ as a function of $\mathbf{A}^d$
does not exist for this formulation in general. However, we want to emphasize that
the non-equilibrated strain energy function $\mathcal{W}_{neq}$ only
represents a deviation from the thermodynamical equilibrium, hence the
possible nonsymmetry of $\mathbb{C}_{neq}^{|d}$ becomes less relevant if the
symmetry of the total tangent moduli $\mathbb{C}=\mathbb{C}_{eq}%
+\mathbb{C}_{neq}$ is assessed. It can be shown that the total tangent moduli
$\mathbb{C}$ results to be numerically\ symmetric for the special case of
isotropic materials undergoing large shear deformations (first example below)
and exactly symmetric for the special case of orthotropic materials undergoing finite
deformations along the preferred material directions (second example below).
Furthermore, for orthotropic materials undergoing large off-axis deformations
and large perturbations away from thermodynamical equilibrium, a very good
convergence rate is still attained during the global finite element iterations
using the symmetric part of Eq. (\ref{C neq 0}) and a symmetric solver (third
example below). In order to symmetrize the tensor $\mathbb{C}_{neq}^{|d}$ of Eq. (\ref{C neq 0}), just substitute
$\mathbf{S}_{neq}^{|tr}\,^{tr}\mathbf{C}_{e}\mathbf{C}^{d-1}$ by its symmetric part, i.e. $\mathbf{S}_{neq}^{|d}$,
see Eq. (\ref{S neq 0}).

The trial tensors $\mathbf{S}_{neq}^{|tr}$ and $\mathbb{C}_{neq}^{|tr}$, present
in Eqs. (\ref{S neq 0}) and (\ref{C neq 0}), may be obtained from our model,
based on logarithmic strains, through%
\begin{equation}
\mathbf{S}_{neq}^{|tr}=\left.  \dfrac{\delta\mathcal{W}_{neq}}{\delta
\,^{tr}\mathbf{A}_{e}}\right\vert _{\xvdo}=\left.  \dfrac{\delta
\mathcal{W}_{neq}}{\delta\,^{tr}\mathbf{E}_{e}}\right\vert _{\xvdo}%
:\dfrac{d\,^{tr}\mathbf{E}_{e}}{d\,^{tr}\mathbf{A}_{e}%
}=:\mathbf{T}_{neq}^{|tr}:\dfrac{d\,^{tr}\mathbf{E}_{e}}%
{d\,^{tr}\mathbf{A}_{e}} \label{S neq trial state}%
\end{equation}
and ---note that $d\,^{tr}\mathbf{E}_{e}/d\,^{tr}\mathbf{A}_{e}$
has major and minor symmetries and that it represents a formal (total) gradient%
\begin{equation}
\mathbb{C}_{neq}^{|tr}=\dfrac{d\mathbf{S}_{neq}^{|tr}}{d\,^{tr}\mathbf{A}_{e}}%
=\dfrac{d\,^{tr}\mathbf{E}_{e}}{d\,^{tr}\mathbf{A}_{e}}%
:\dfrac{d\mathbf{T}_{neq}^{|tr}}{d\,^{tr}\mathbf{E}_{e}%
}:\dfrac{d\,^{tr}\mathbf{E}_{e}}{d\,^{tr}\mathbf{A}_{e}%
}+\mathbf{T}_{neq}^{|tr}:\dfrac{d^{2}\,^{tr}\mathbf{E}_{e}}%
{d\,^{tr}\mathbf{A}_{e}d\,^{tr}\mathbf{A}_{e}}
\label{C neq trial state}%
\end{equation}
The trial generalized Kirchhoff stress tensor $\mathbf{T}_{neq}^{|tr}$
has to be previously related to the updated generalized Kirchhoff stress tensor
$\mathbf{T}_{neq}^{|e}$, which is the resulting stress tensor at each global
iteration obtained from $\mathcal{W}_{neq}\left(  \mathbf{E}_{e}^d\right)  $
using Eq. (\ref{T hat neq}). We have
---compare to\ Eq. (\ref{SPK stresses neq}) and consider the change of
variable $\mathbf{E}$ by $\,^{tr}\mathbf{E}_{e}$%
\begin{equation}
\mathbf{T}_{neq}^{|tr}=\left.  \dfrac{\delta\mathcal{W}_{neq}}{\delta
\,^{tr}\mathbf{E}_{e}}\right\vert _{\xvdo}=\dfrac{d\mathcal{W}_{neq}%
}{d\mathbf{E}_{e}}:\left.  \dfrac{\delta\mathbf{E}_{e}}{\delta
\,^{tr}\mathbf{E}_{e}}\right\vert _{\xvdo}=\mathbf{T}_{neq}^{|e}:\left.
\dfrac{\delta\mathbf{E}_{e}}{\delta\,^{tr}\mathbf{E}_{e}}\right\vert _{\xvdo}
\label{T neq trial state}%
\end{equation}
Hereafter, analogously as we did in Ref. \cite{LatMonCM2015}, we approximate%
\begin{equation}
\left.  \dfrac{\delta\mathbf{E}_{e}}{\delta\,^{tr}\mathbf{E}_{e}}\right\vert
_{\xvdo}\approx\mathbb{I}^{S}\quad\Rightarrow\quad\mathbf{T}%
_{neq}^{|tr}\approx\mathbf{T}_{neq}^{|e} \label{mapping tensor Ee Eetr}%
\end{equation}
which is an approximation valid for $\Delta t/\tau\ll1$ in the most general
case (as for the model based on the Sidoroff decomposition, note that
$\mathbf{T}_{neq}^{|tr}=\mathbf{T}_{neq}^{|e}$ for the special cases of
isotropic materials under arbitrary loadings or orthotropic materials
undergoing finite deformations along the preferred material directions). If we
do not wish to take this approximation, we should compute the analytical
mapping tensor present in Eq. (\ref{T neq trial state}) and its derivatives in
the numerical algorithm, cf. Ref. \cite{LatMonCM2015}, Appendix $2$.
The modified second Piola--Kirchhoff stresses $\mathbf{S}_{neq}^{|d}$
are obtained combining, first, Eqs. (\ref{S neq trial state}),
(\ref{T neq trial state}) and (\ref{mapping tensor Ee Eetr})$_{1}$
\begin{equation}\label{S neq trial state approx}
\mathbf{S}_{neq}^{|tr}=\left.\dfrac{d\mathcal{W}_{neq}}{d\mathbf{E}%
_{e}}\right\vert_{t+\Delta t}:\dfrac{d\,^{tr}\mathbf{E}_e}{d\,^{tr}\mathbf{A}_e}
\end{equation}
and then performing the mapping from the internal (trial) to the external (isochoric)
configurations using Eq. (\ref{S neq 0}).

In order to obtain the consistent tangent moduli $d\mathbf{T%
}_{neq}^{|tr}/d\,^{tr}\mathbf{E}_{e}$, needed in Eq. (\ref{C neq trial state}%
), we have to take into consideration that the trial logarithmic strains
$\,^{tr}\mathbf{E}_{e}$ and the updated logarithmic strains $\,_{\hspace
{3.2ex}0}^{t+\Delta t}\mathbf{E}_{e}$ are related in the algorithm through Eq.
(\ref{viscous correction}). Hence%
\begin{equation}
\dfrac{d\mathbf{T}_{neq}^{|tr}}{d\,^{tr}\mathbf{E}_{e}}%
=\dfrac{d\mathbf{T}_{neq}^{|e}}{d\,^{tr}\mathbf{E}_{e}}%
=\dfrac{d\mathbf{T}_{neq}^{|e}}{d\mathbf{E}_{e}}:\dfrac
{d\,_{\hspace{3.2ex}0}^{t+\Delta t}\mathbf{E}_{e}}{d
\,^{tr}\mathbf{E}_{e}} \label{T/E neq trial state}%
\end{equation}
with the tensor $d\,_{\hspace{3.2ex}0}^{t+\Delta t}\mathbf{E}%
_{e}/d\,^{tr}\mathbf{E}_{e}$ providing the consistent linearization of
the algorithmic formulation during the viscous correction substep. Taking
derivatives in Eq. (\ref{viscous correction}), we identify
\begin{equation}
\dfrac{d\mathbf{T}_{neq}^{|tr}}{d\,^{tr}\mathbf{E}_{e}%
}=\left.  \dfrac{d^{2}\mathcal{W}_{neq}}{d\mathbf{E}_{e}%
d\mathbf{E}_{e}}\right\vert _{t+\Delta t}:\left.  \frac{d
\mathbf{R_E}}{d\mathbf{E}_{e}}\right\vert _{t+\Delta t}^{-1}
\label{algorithmic tangent}%
\end{equation}
where the algorithmic gradient $d\,_{\hspace{3.2ex}0}^{t+\Delta
t}\mathbf{E}_{e}/d\,^{tr}\mathbf{E}_{e}$ is given by the inverse of Eq.
(\ref{dR/dEe}) evaluated at the updated strains $\,_{\hspace{3.2ex}%
0}^{t+\Delta t}\mathbf{E}_{e}$, see Section
\ref{Section - Local Newton iterations for the non-equilibrated part}. Note
that only the deviatoric part of this tensor is relevant in Eq.
(\ref{algorithmic tangent}). Interestingly, although the algorithmic gradient $d
\,_{\hspace{3.2ex}0}^{t+\Delta t}\mathbf{E}_{e}/d\,^{tr}\mathbf{E}_{e}$
is, in general, non-symmetric in this case, the trial consistent tangent tensor
$d\mathbf{T}_{neq}^{|tr}/d\,^{tr}\mathbf{E}_{e}$, as given in
Eqs. (\ref{T/E neq trial state}) or (\ref{algorithmic tangent}), has major and
minor symmetries. This is thanks to the fact that the viscosity tensor $\mathbb{V}%
^{-1}$ in Eq. (\ref{viscous flow rule}) is fully symmetric
\cite{ReeseGovindjee,LatMonCM2015}.
The modified consistent (non-symmetric, in general) tangent moduli
$\mathbb{C}_{neq}^{|d}$ for the non-equilibrated part is obtained combining, first,
Eqs. (\ref{C neq trial state}), (\ref{mapping tensor Ee Eetr})$_{2}$ and
(\ref{algorithmic tangent}) ---note that $\mathbb{C}_{neq}^{|tr}$ preserves
major and minor symmetries%
\begin{align}
\mathbb{C}_{neq}^{|tr}&=\dfrac{d\,^{tr}\mathbf{E}_{e}}{d
\,^{tr}\mathbf{A}_{e}}:\left.  \dfrac{d^{2}\mathcal{W}_{neq}}%
{d\mathbf{E}_{e}d\mathbf{E}_{e}}\right\vert _{t+\Delta t}%
:\dfrac{d\,_{\hspace{3.5ex}0}^{t+\Delta t}\mathbf{E}_{e}}%
{d\,^{tr}\mathbf{E}_{e}}:\dfrac{d\,^{tr}\mathbf{E}_{e}}%
{d\,^{tr}\mathbf{A}_{e}}\nonumber\\
&\ \ +\left.  \dfrac{d\mathcal{W}_{neq}%
}{d\mathbf{E}_{e}}\right\vert _{t+\Delta t}:\dfrac{d^{2}%
\,^{tr}\mathbf{E}_{e}}{d\,^{tr}\mathbf{A}_{e}d\,^{tr}%
\mathbf{A}_{e}}\label{C neq trial state approx}%
\end{align}
and then mapping the result from the internal to the external configurations using Eq.
(\ref{C neq 0}). Mapping tensors relating material logarithmic strains to Green--Lagrange strains
are given in spectral form in, for example, Ref. \cite{LatMonCM2014}, Section $2.5$.

Finally, the isochoric non-equilibrated stresses $\mathbf{S}_{neq}$ and consistent tangent moduli $\mathbb{C}_{neq}=d\mathbf{S}_{neq}/d\mathbf{A}$ are
obtained from $\mathbf{S}_{neq}^{|d}$ and $\mathbb{C}_{neq}^{|d}=d\mathbf{S}_{neq}^{|d}/d\mathbf{A}^d$ using the
deviatoric projection tensor $d\mathbf{A}^d/d\mathbf{A}$ (recall Eq. (\ref{S_neq_dev})) and its derivatives through ---see Ref. \cite{LatMonCM2015}, Appendix \emph{1}
\begin{equation}\label{S neq projection}
\mathbf{S}_{neq}=J^{-2/3}\mathbf{S}_{neq}^{|d}
\end{equation}
and
\begin{equation}\label{C neq projection}
\mathbb{C}_{neq}=J^{-4/3}\mathbb{C}_{neq}^{|d}
\end{equation}
In the derivation of Eq. (\ref{C neq projection}) we have used the fact that the second
and third addends in the right-hand side of Eq. (\ref{C neq 0}) cancel to each other when the
two-index contraction operations $\mathbf{C}^d:\mathbb{C}_{neq}^{|d}$ and $\mathbb{C}_{neq}^{|d}:\mathbf{C}^d$ are performed,
which allows us to consider the symmetry relation $\mathbf{C}^d:\mathbb{C}_{neq}^{|d}=\mathbb{C}_{neq}^{|d}:\mathbf{C}^d$.

\subsection{Linearized case: Finite linear
viscoelasticity\label{Linearized case: Finite linear viscoelasticity}}

The constitutive equation for the viscous flow Eq. (\ref{viscous flow rule})
may be simplified when either linear finite logarithmic or linear small
stress-strain relations are derived from the non-equilibrated contribution
$\mathcal{W}_{neq}$. In both cases, the same linear/linearized solution for
the evolution equation is obtained, i.e. the so-called Finite Linear
Viscoelasticity. The (linear) viscous flow rule for the fully orthotropic
model using the reverse multiplicative decomposition represents a
generalization of the expression derived for the Sidoroff's decomposition
---c.f. Ref. \cite{LatMonCM2015}%

\begin{equation}
-\left.  \dfrac{d\mathbf{E}_{e}}{dt}\right\vert _{\edo}=\mathbb{P}^{S}:\left(
\mathbb{\bar{V}}^{-1}:\mathbb{P}^{S}:\dfrac{d^{2}\mathcal{W}_{neq}%
}{d\mathbf{E}_{e}^{d}d\mathbf{E}_{e}^{d}}\right)  :\mathbb{P}%
^{S}:\mathbf{E}_{e}=\mathbb{T}^{-1}_d:\mathbf{E}_{e}
\label{viscous flow rule linear}%
\end{equation}
where $\mathbf{E}_{e}$ is used in this section to represent either the
internal elastic logarithmic strain tensor or the internal
elastic\ infinitesimal strains tensor $\mathbf{\varepsilon}_{e}$. The
simplification in this case emerges from the fact that the integration of Eq.
(\ref{viscous flow rule linear}) during the viscous corrector substep gives an
explicit update for $\,_{\hspace{3.2ex}0}^{t+\Delta t}\mathbf{E}_{e}$ in terms
of $\,^{tr}\mathbf{E}_{e}$, i.e. ---compare to Eq. (\ref{viscous correction})
for the fully non-linear case%
\begin{equation}
\left(  \mathbb{I}^{S}+\Delta t\mathbb{T}^{-1}_d\right)  :\,_{\hspace
{3.2ex}0}^{t+\Delta t}\mathbf{E}_{e}=\,^{tr}\mathbf{E}_{e}\quad\Rightarrow
\quad\,_{\hspace{3.2ex}0}^{t+\Delta t}\mathbf{E}_{e}=\left(  \mathbb{I}%
^{S}+\Delta t\mathbb{T}^{-1}_d\right)  ^{-1}:\,^{tr}\mathbf{E}_{e}%
\end{equation}
so no local Newton iterations are needed.

\section{Equilibrated
contribution\label{Section - Deviatoric and volumetric equilibrated contributions to stress and tangent}
}

If the total gradient $\,_{\hspace{3.2ex}0}^{t+\Delta t}\mathbf{X}$ is known
at time step $t+\Delta t$, then the equilibrated contributions
$\,^{t+\Delta t}\mathbf{S}_{eq}$ and $\,^{t+\Delta t}\mathbb{C}_{eq}$ are just
obtained from $\Psi_{eq}(\mathbf{E})=\mathcal{W}_{eq}(\mathbf{E}%
^{d})+\mathcal{U}_{eq}(J)$ as usual uncoupled deviatoric-volumetric hyperelastic calculations, i.e.%
\begin{equation}
\mathbf{S}_{eq}=\dfrac{d\Psi_{eq}}{d\mathbf{A}}=\dfrac
{d\Psi_{eq}}{d\mathbf{E}}:\dfrac{d\mathbf{E}}%
{d\mathbf{A}}=\mathbf{T}_{eq}:\dfrac{d\mathbf{E}}{d
\mathbf{A}} \label{S eq}%
\end{equation}%
\begin{equation}
\mathbb{C}_{eq}=\dfrac{d\mathbf{S}_{eq}}{d\mathbf{A}}%
=\dfrac{d\mathbf{E}}{d\mathbf{A}}:\dfrac{d\mathbf{T}%
_{eq}}{d\mathbf{E}}:\dfrac{d\mathbf{E}}{d\mathbf{A}%
}+\mathbf{T}_{eq}:\dfrac{d^{2}\mathbf{E}}{d\mathbf{A}%
d\mathbf{A}} \label{C eq}%
\end{equation}
For detailed formulae to compute these
contributions for an incompressible orthotropic material, we refer to Ref.
\cite{LatMonCM2014}, Section $2.5$.

\section{Determination of the viscosity parameters of the orthotropic
model\label{Section - Determination of the relaxation time(s) of the orthotropic model}%
}

Consider a small strains uniaxial relaxation test performed about the
preferred material direction $\mathbf{e}_{1}$ of an incompressible material.
Equation (\ref{viscous flow rule linearized}) represented in preferred
material axes and specialized at $t=0^{+}$ (just after the total deformation
in direction $\mathbf{e}_{1}$ is applied and retained) reads ---note that
shear terms are not needed and that $\mathbf{\varepsilon}_{e}^{0}%
=\mathbf{\varepsilon}_{e}(t=0^{+})=\mathbf{\varepsilon}(t=0^{+}%
)=\mathbf{\varepsilon}^{0}$ are isochoric (traceless)%

\begin{equation}
-\left[
\begin{array}
[c]{c}%
\dot{\varepsilon}_{e11}^{0}\smallskip\\
\dot{\varepsilon}_{e22}^{0}\smallskip\\
\dot{\varepsilon}_{e33}^{0}%
\end{array}
\right]  _{\scriptsize{\mathbf{\dot{\varepsilon}}=\mathbf{0}}}=\frac{\varepsilon_{11}^{0}}{9}\left[
\begin{array}
[c]{c}%
2\dfrac{E_{11}^{neq}}{\eta_{11}^{d}}+\dfrac{H_{12}^{neq}}{\eta_{22}^{d}%
}+\dfrac{H_{13}^{neq}}{\eta_{33}^{d}}\smallskip\\
-\dfrac{E_{11}^{neq}}{\eta_{11}^{d}}-2\dfrac{H_{12}^{neq}}{\eta_{22}^{d}%
}+\dfrac{H_{13}^{neq}}{\eta_{33}^{d}}\smallskip\\
-\dfrac{E_{11}^{neq}}{\eta_{11}^{d}}+\dfrac{H_{12}^{neq}}{\eta_{22}^{d}%
}-2\dfrac{H_{13}^{neq}}{\eta_{33}^{d}}%
\end{array}
\right]  \label{epse0 dot}%
\end{equation}
where we have defined%
\begin{equation}%
\begin{array}
[c]{c}%
E_{11}^{neq}:=2\mu_{11}^{neq}+\mu_{22}^{neq}\nu_{12}^{0}+\mu_{33}^{neq}%
\nu_{13}^{0}\smallskip\\
H_{12}^{neq}:=\mu_{11}^{neq}+2\mu_{22}^{neq}\nu_{12}^{0}-\mu_{33}^{neq}%
\nu_{13}^{0}\smallskip\\
H_{13}^{neq}:=\mu_{11}^{neq}-\mu_{22}^{neq}\nu_{12}^{0}+2\mu_{33}^{neq}%
\nu_{13}^{0}%
\end{array}
\label{Hs def}%
\end{equation}
and the initial Poisson ratios $\nu_{12}^{0}:=-\varepsilon_{22}^{0}%
/\varepsilon_{11}^{0}$ and $\nu_{13}^{0}:=-\varepsilon_{33}^{0}/\varepsilon
_{11}^{0}$ are expressed in terms of the equilibrated and
non-equilibrated\ reference shear moduli through ---c.f. Ref.
\cite{LatMonCM2015}%
\begin{align}
\nu_{12}^{0}  &  =\frac{\mu_{33}^{0}}{\mu_{22}^{0}+\mu_{33}^{0}}=\frac
{\mu_{33}^{eq}+\mu_{33}^{neq}}{\mu_{22}^{eq}+\mu_{22}^{neq}+\mu_{33}^{eq}%
+\mu_{33}^{neq}}\smallskip\label{nu0 12}\\
\nu_{13}^{0}  &  =\frac{\mu_{22}^{0}}{\mu_{22}^{0}+\mu_{33}^{0}}=\frac
{\mu_{22}^{eq}+\mu_{22}^{neq}}{\mu_{22}^{eq}+\mu_{22}^{neq}+\mu_{33}^{eq}%
+\mu_{33}^{neq}} \label{nu0 13}%
\end{align}
The uniaxial stress at $t=0^{+}$ is found to be%
\begin{equation}
\sigma_{11}^{0}=\left(  2\mu_{11}^{0}+\mu_{22}^{0}\nu_{12}^{0}+\mu_{33}^{0}%
\nu_{13}^{0}\right)  \varepsilon_{11}^{0}=:E_{11}^{0}\varepsilon_{11}^{0}
\label{sig011 eps011}%
\end{equation}
where $E_{11}^{0}$ represents the instantaneous Young's modulus in direction
$\mathbf{e}_{1}$. Upon the split
\begin{equation}
\sigma_{11}^{0}=E_{11}^{eq}\varepsilon_{11}^{0}+E_{11}^{neq}\varepsilon
_{e11}^{0}%
\end{equation}
with%
\begin{equation}
E_{11}^{eq}:=2\mu_{11}^{eq}+\mu_{22}^{eq}\nu_{12}^{0}+\mu_{33}^{eq}\nu
_{13}^{0}%
\end{equation}
the consideration of the first component in Eq. (\ref{epse0 dot}) and the
subsequent comparison of Eq. (\ref{sig011 eps011}) to the expression of its
time derivative $\dot{\sigma}^{0}_{11}$ we readily arrive to%
\begin{equation}
\left\{
\begin{array}
[c]{c}%
2\dfrac{E_{11}^{neq}}{\mu_{11}^{neq}}\dfrac{1}{\tau_{11}}+\dfrac{H_{12}^{neq}%
}{\mu_{22}^{neq}}\dfrac{1}{\tau_{22}}+\dfrac{H_{13}^{neq}}{\mu_{33}^{neq}%
}\dfrac{1}{\tau_{33}}=\dfrac{9}{t_{11}^{0}}\left(  1+\dfrac{E_{11}^{eq}%
}{E_{11}^{neq}}\right)  \bigskip\\
\dfrac{H_{21}^{neq}}{\mu_{11}^{neq}}\dfrac{1}{\tau_{11}}+2\dfrac{E_{22}^{neq}%
}{\mu_{22}^{neq}}\dfrac{1}{\tau_{22}}+\dfrac{H_{23}^{neq}}{\mu_{33}^{neq}%
}\dfrac{1}{\tau_{33}}=\dfrac{9}{t_{22}^{0}}\left(  1+\dfrac{E_{22}^{eq}%
}{E_{22}^{neq}}\right)  \bigskip\\
\dfrac{H_{31}^{neq}}{\mu_{11}^{neq}}\dfrac{1}{\tau_{11}}+\dfrac{H_{32}^{neq}%
}{\mu_{22}^{neq}}\dfrac{1}{\tau_{22}}+2\dfrac{E_{33}^{neq}}{\mu_{33}^{neq}%
}\dfrac{1}{\tau_{33}}=\dfrac{9}{t_{33}^{0}}\left(  1+\dfrac{E_{33}^{eq}%
}{E_{33}^{neq}}\right)  \bigskip
\end{array}
\right.  \label{tau 11 test}%
\end{equation}
where the experimental value $t_{11}^{0}:=-\sigma_{11}^{0}/\dot{\sigma}%
_{11}^{0}$ may be measured tracing the tangent to the relaxation curve
$\sigma_{11}(t)$ at $t=0^{+}$. In Equations (\ref{tau 11 test}) the values of
$H_{ij}$ and $E_{ii}^{neq}$ are defined as in Eq. (\ref{Hs def}) for their
respective direction. Equations (\ref{tau 11 test}) are linear in the three
independent (inverse) relaxation times $1/\tau_{11}$, $1/\tau_{22}$ and
$1/\tau_{33}$. Hence, they constitute a linear system from which we can
determine the relaxation times $\tau_{11}$, $\tau_{22}$ and $\tau_{33}$ once
the experimental factors $t_{11}^{0}$, $t_{22}^{0}$ and $t_{33}^{0}$ are known
(measured). The three viscosities $\eta_{11}^{d}$, $\eta_{22}^{d}$ and
$\eta_{33}^{d}$ are then obtained from Eq. (\ref{relaxation times 0}). If the
three \textquotedblleft axial\textquotedblright\ viscosities are equal, i.e.
$\mu_{11}^{neq}\tau_{11}=\mu_{22}^{neq}\tau_{22}=\mu_{33}^{neq}\tau_{33}%
=\eta^{d}$, then Eq. (\ref{tau 11 test})$_1$ is uncoupled from the other preferred
directions and provides the same result obtained in Ref. \cite{LatMonCM2015}%
\begin{equation}
\dfrac{\eta^{d}}{\mu_{11}^{neq}}=\tau_{11}=t_{11}^{0}\frac{E_{11}^{neq}%
/(3\mu_{11}^{neq})}{1+E_{11}^{eq}/E_{11}^{neq}}%
\end{equation}

For the present orthotropic case, three additional small strain shear tests
are needed in order to completely characterize the present model.
Since the shear components in Eq. (\ref{viscous flow rule linearized}) are
fully uncoupled in the preferred material basis, we obtain from a simple
shear relaxation test in the plane $\{\mathbf{e}_{1},\mathbf{e}_{2}\}$ under a
plane stress condition %
\begin{equation}
\frac{\eta_{12}^{d}}{\mu_{12}^{neq}}=\tau_{12}=t_{12}^{0}\frac{1}{1+\mu
_{12}^{eq}/\mu_{12}^{neq}}%
\end{equation}
where $t_{12}^{0}:=-\sigma_{12}^{0}/\dot{\sigma}_{12}^{0}$ is determined from
the experimental shear stress relaxation curve $\sigma_{12}(t)$. Two homologous
expressions for $\tau_{23}$ and $\tau_{31}$ are derived from the respective
simple shear tests performed in the other preferred planes $\{\mathbf{e}%
_{2},\mathbf{e}_{3}\}$ and $\{\mathbf{e}_{3},\mathbf{e}_{1}\}$.

\section{Examples}

The following examples are designed to compare the obtained behavior against
models based on the Sidoroff decomposition \cite{ReeseGovindjee,LatMonCM2015}
and to highlight the enhanced capabilities of the present
anisotropic visco-hyperelasticity formulation based on the reverse multiplicative decomposition.

\subsection{Isotropic material}

\begin{figure}
[ptb]
\begin{center}
\includegraphics[width=0.9\textwidth]%
{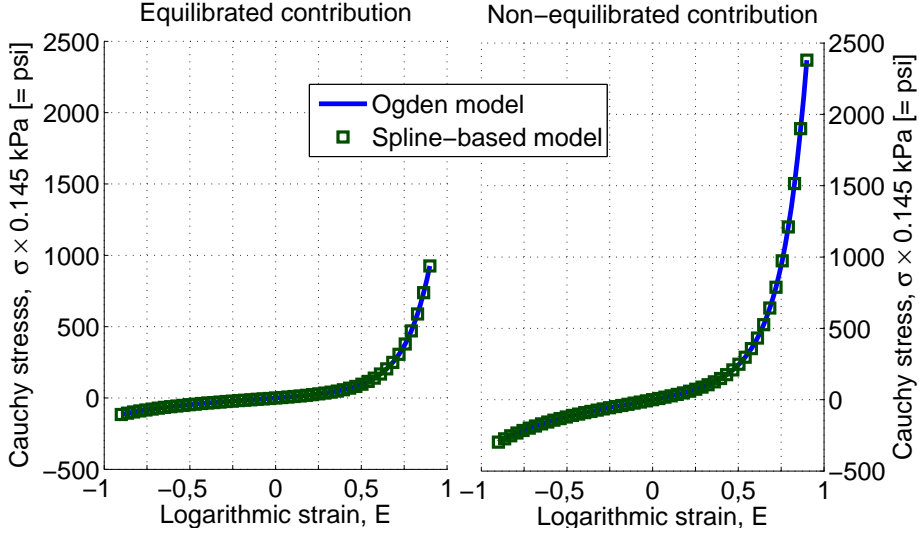}%
\caption{Uniaxial stresses derived from the equilibrated and non-equilibrated
Ogden-type strain energy functions \cite{ReeseGovindjee} and predictions using
the spline-based isotropic model \cite{SusBat2009}.}%
\label{figure05.eps}%
\end{center}
\end{figure}

In Ref. \cite{LatMonCM2015} we showed that the anisotropic model derived
therein in full material description gives exactly the same results than the
isotropic model of Reese and Govindjee \cite{ReeseGovindjee} formulated in
spatial principal directions when an isotropic (spline-based, Ogden-type or
whatever) strain energy function is considered. Both models are based on the
Sidoroff's multiplicative decomposition. The same simple shear cyclic test
simulation of the first example in Ref. \cite{LatMonCM2015} is performed
herein, the only difference between them being the multiplicative
decomposition employed in the formulation. That is, the same equilibrated and
non-equilibrated deviatoric strain energy functions $\mathcal{W}_{eq}$ and
$\mathcal{W}_{neq}$ (using the spline-based model, see Figure
\ref{figure05.eps}), volumetric penalty function $\mathcal{U}_{eq}$,
relaxation time $\tau=\eta^{d}/\mu^{neq}=17.5%
\operatorname{s}%
$ and (mixed) finite element formulation are employed. We only compare the
results obtained with the respective Finite (Non-Linear) Viscoelasticity
formulations.

\begin{figure}
[ptb]
\begin{center}
\includegraphics[width=1.\textwidth]%
{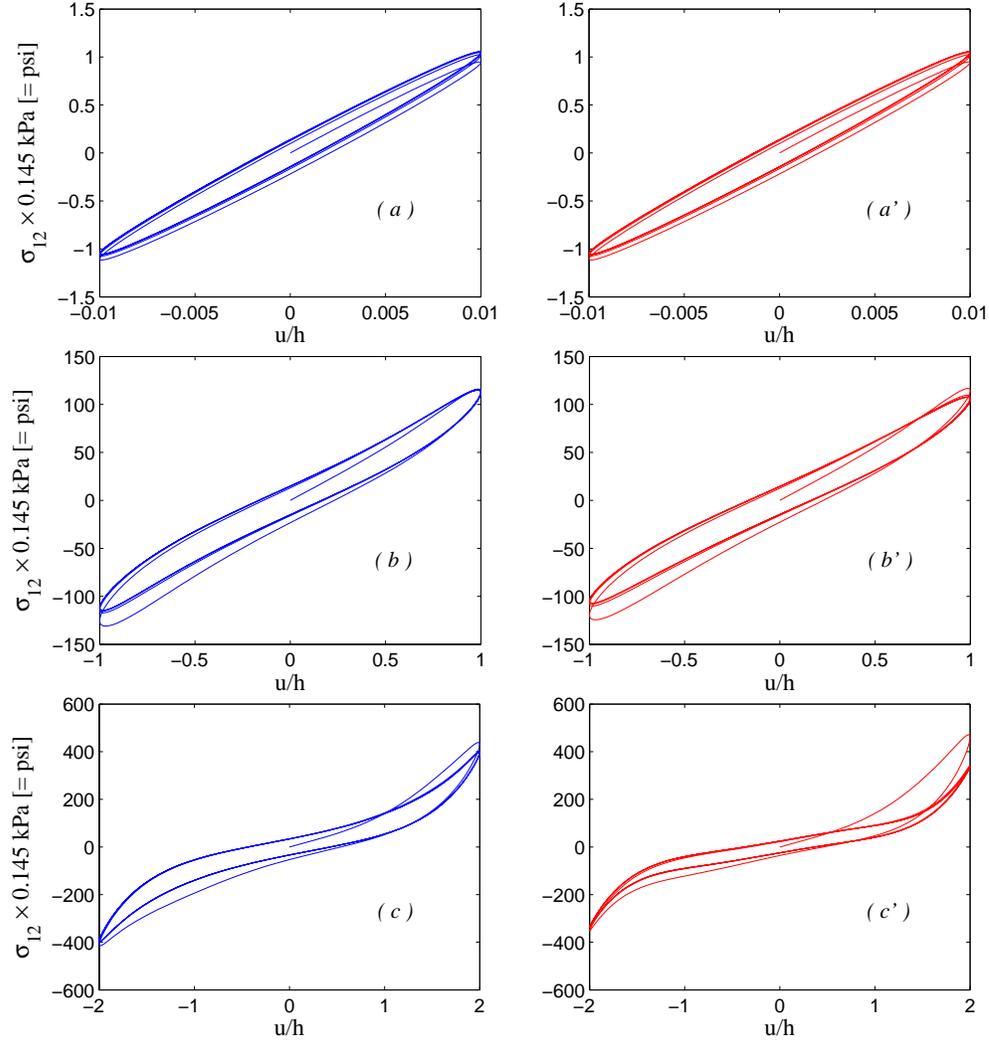}%
\caption{Cauchy shear stresses $\sigma_{12}(t)$ versus engineering shear
strains $\gamma_{12}(t)=u(t)/h=u_{0}/h\times sin(0.3t)$ for the amplitudes: a)
a') $u_{0}/h=0.01$, b) b') $u_{0}/h=1$, c) c') $u_{0}/h=2$. Curves \emph{a},
\emph{b} and \emph{c} obtained using the model based on the Sidoroff
decomposition as given in Ref. \cite{LatMonCM2015} (see also
\cite{ReeseGovindjee}). Curves \emph{a'}, \emph{b'} and \emph{c'} are obtained
using the present formulation based on the reversed decomposition. All the
simulations are performed using the respective finite fully non-linear
formulations (FV) (100 time steps per cycle).}%
\label{figure06.eps}%
\end{center}
\end{figure}

In Figure \ref{figure06.eps}, the obtained Cauchy shear stresses $\sigma
_{12}(t)$ are plotted against the engineering shear strains $\gamma_{12}(t)$
for three different amplitudes in the simple shear test.%

Both models predict the same behavior within the context of small strains. However, the predictions given by the models separate when large shear strains are
considered (note that for purely axial loadings, both models would exactly predict the
same viscoelastic behavior).

Representative convergence rates for the unbalanced force and energy using a
symmetric solver and an unsymmetric solver are shown in Table \ref{Tabla 1} for
the case labeled $(c^{\prime})$ in Figure \ref{figure06.eps}. The symmetrization of
the third addend in Eq. (\ref{C neq 0}) and the subsequent use of a symmetric
solver in the global finite element iterations are clearly justified in this case.

\begin{table}[h]%
\begin{tabular}
[c]{|lllll|}\hline%
\begin{tabular}
[c]{@{}l}%
Step 465\\
Iteration
\end{tabular}
&
\begin{tabular}
[c]{@{}l}%
Load norm\\
(Symmetric)
\end{tabular}
&
\begin{tabular}
[c]{@{}l}%
Load norm\\
(Unsymmetric)
\end{tabular}
&
\begin{tabular}
[c]{@{}l}%
Energy norm\\
(Symmetric)
\end{tabular}
&
\begin{tabular}
[c]{@{}l}%
Energy norm\\
(Unsymmetric)
\end{tabular}
\\\hline\hline
\multicolumn{1}{|c}{1} & \multicolumn{1}{l}{1.412E+05} &
\multicolumn{1}{l}{1.412E+05} & \multicolumn{1}{l}{1.771E+04} &
\multicolumn{1}{l|}{1.771E+04}\\
\multicolumn{1}{|c}{2} & \multicolumn{1}{l}{7.923E+00} &
\multicolumn{1}{l}{5.307E+00} & \multicolumn{1}{l}{5.455E-05} &
\multicolumn{1}{l|}{1.862E-06}\\
\multicolumn{1}{|c}{3} & \multicolumn{1}{l}{3.334E-03} &
\multicolumn{1}{l}{1.692E-04} & \multicolumn{1}{l}{6.050E-11} &
\multicolumn{1}{l|}{9.427E-12}\\\hline
\end{tabular}
\caption{Comparison of convergence rates using the reverse decomposition and
either a symmetric solver or an unsymmetric solver. Case labeled $(c^{\prime})$ in Figure \ref{figure06.eps}}%
\label{Tabla 1}%
\end{table}

\subsection{Orthotropic material with linear logarithmic stress-strain
relations}

In this example from Reference \cite{LatMonCM2015} uniaxial in-axis
orthotropic relaxation testing is performed along different material directions. Consider the following strain energy
functions%
\begin{equation}
\mathcal{W}_{eq}(\mathbf{E}^{d})=\mu_{11}^{eq}(E_{11}^{d})^{2}+\mu_{22}%
^{eq}(E_{22}^{d})^{2}+\mu_{33}^{eq}(E_{33}^{d})^{2} \label{Weq axial}%
\end{equation}%
\begin{equation}
\mathcal{W}_{neq}(\mathbf{E}_{e}^{d})=\mu_{11}^{neq}(E_{e11}^{d})^{2}+\mu
_{22}^{neq}(E_{e22}^{d})^{2}+\mu_{33}^{neq}(E_{e33}^{d})^{2}
\label{Wneq axial}%
\end{equation}
where only the axial components in principal material directions are needed
in order to simulate the different uniaxial relaxation tests about the
preferred material axes. We take the same values for the shear moduli in Eqs.
(\ref{Weq axial}) and (\ref{Wneq axial}) used in the second example in Ref.
\cite{LatMonCM2015}%
\begin{equation}
\mu_{11}^{eq}=4%
\operatorname{MPa}%
\text{, }\mu_{22}^{eq}=2%
\operatorname{MPa}%
\text{, }\mu_{33}^{eq}=1%
\operatorname{MPa}
\label{deviatoric moduli eq}%
\end{equation}%
\begin{equation}
\mu_{11}^{neq}=5%
\operatorname{MPa}%
\text{, }\mu_{22}^{neq}=3%
\operatorname{MPa}%
\text{, }\mu_{33}^{neq}=2%
\operatorname{MPa}
\label{deviatoric moduli neq}%
\end{equation}
In the example of Ref. \cite{LatMonCM2015}, a single relaxation time
$\tau_{11}=20%
\operatorname{s}%
$ was needed in order to complete the definition of the model. The
(non-independent) relaxation times $\tau_{22}=\tau_{11}\times\mu_{11}%
^{neq}/\mu_{22}^{neq}=33.3%
\operatorname{s}%
$ and $\tau_{33}=\tau_{11}\times\mu_{11}^{neq}/\mu_{33}^{neq}=50%
\operatorname{s}%
$ were then obtained\ by the model because of the isotropy assumption in the
viscous component. In the present anisotropic case we need to prescribe three
independent relaxation times (the three other relaxation times for shear
behavior are not needed in this example). The initially undeformed block of
$100\times100\times100$ is deformed (quasi) instantaneously along material
direction $1$ up to a dimension of $300$, whereas the other directions, due to
material behavior, result in $66.2$ for material direction $2$ and $50.4$ for
material direction $3$; see Ref. \cite{LatMonCM2015}.

\begin{figure}
[ptb]
\begin{center}
\includegraphics[width=0.6\textwidth]%
{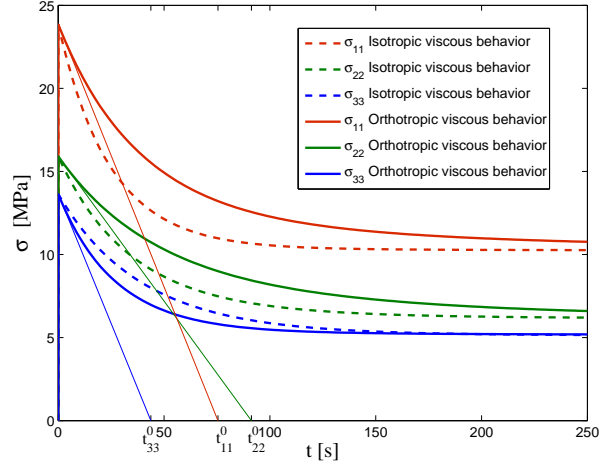}%
\caption{Dashed curves: Stress relaxation curves $\sigma_{11}(t)$,
$\sigma_{22}(t)$ and $\sigma_{33}(t)$ obtained from three uniaxial relaxation
tests performed about the preferred material directions $1$, $2$ and $3$,
respectively, using the model based on the reverse multiplicative
decomposition and the same isotropic viscosity tensor used in the second
example of Ref. \cite{LatMonCM2015}. Solid curves: idem using an orthotropic
viscosity tensor.}%
\label{figure08.eps}%
\end{center}
\end{figure}

In the first simulation within this example we prescribe the relaxation times
that give as a result the same isotropic viscosity tensor used in Ref.
\cite{LatMonCM2015} ---see Eq. (\ref{relaxation times 0})%
\begin{equation}
\tau_{11}=20%
\operatorname{s}%
~,\quad\tau_{22}=33.3%
\operatorname{s}%
~,\quad\tau_{33}=50%
\operatorname{s}%
\quad\Rightarrow\quad\eta_{11}^{d}=\eta_{22}^{d}=\eta_{33}^{d}
\label{relaxation times isotropic}%
\end{equation}
in order to show that the same results are obtained using the model based on
the Sidoroff multiplicative decomposition and the present model based on the
reversed one, both with the same isotropic viscous behavior. The same results
are expected to be obtained because the loads are applied over the preferred
directions of the material and no rotations are present. Hence both
decompositions are indistinguishable from a numerical standpoint, i.e.
$\mathbf{U}=\mathbf{U}_{e}\mathbf{U}_{v}=\mathbf{U}_{v}\mathbf{U}_{e}$.
Furthermore, in this case the condition for the co-rotational rate
$\mathbf{\mathring{X}}_{v}=\mathbf{0}$ is coincident to the condition
$\mathbf{\dot{X}}_{v}=\mathbf{0}$, whereupon the non-equilibrated tangent
moduli $\mathbb{C}_{neq}$ given in Eq. (\ref{C neq 0}) preserves all the
symmetries and no distinction between using a symmetric or a unsymmetric
solver is needed in this example. In Figure \ref{figure08.eps} we can verify
that identical stress relaxation curves (dashed lines) to those shown in Ref.
\cite{LatMonCM2015} are obtained for the three (separate) uniaxial tests
performed over the three preferred directions using the present model.

As a second case within this example we prescribe independent relaxation times
in order to show the enhanced capabilities of the present model when it is
used with an orthotropic viscosity tensor. The following independently
user-prescribed values for the \emph{axial} relaxation times have been chosen%
\begin{equation}
\tau_{11}=80%
\operatorname{s}%
~,\quad\tau_{22}=100%
\operatorname{s}%
~,\quad\tau_{33}=25%
\operatorname{s}%
\quad\Rightarrow\quad\eta_{11}^{d}\neq\eta_{22}^{d}\neq\eta_{33}^{d}\neq
\eta_{11}^{d} \label{relaxation times orthotropic}%
\end{equation}
In Eq. (\ref{relaxation times isotropic}) the relations $\tau_{11}<\tau
_{22}<\tau_{33}$ hold because $\mu_{11}^{neq}>\mu_{22}^{neq}>\mu_{33}^{neq}$,
i.e. the stiffer non-equilibrated behavior in a given direction, the faster
relaxation process associated to that direction. However, these restrictions
do not necessarily hold in the present model, see Eq.
(\ref{relaxation times orthotropic}). Indeed, we can observe in Figure
\ref{figure08.eps} (solid lines) that in this case the material relaxes faster
in direction $3$ than in the other two directions.%
Since the same strain energy functions are used in all the simulations, the
same instantaneous and relaxed states are obtained for each test,
independently of the relaxation times being prescribed.

Finally, introducing the material parameters given in Eqs.
(\ref{deviatoric moduli eq}), (\ref{deviatoric moduli neq}) and
(\ref{relaxation times orthotropic}) into Eqs. (\ref{tau 11 test}) we obtain
the following values for the experimental parameters%
\begin{equation}
t_{11}^{0}=75.5%
\operatorname{s}%
\text{,\quad}t_{22}^{0}=91.3%
\operatorname{s}%
\text{\quad and\quad}t_{33}^{0}=43.8%
\operatorname{s}%
\end{equation}
We can see in Figure \ref{figure08.eps} that the values $t_{11}^{0}$,
$t_{22}^{0}$ and $t_{33}^{0}$ obtained from the computational relaxation
curves are in very good agreement with the preceding values. This fact proves
the applicability of the material characterization procedure explained in
Section
\ref{Section - Determination of the relaxation time(s) of the orthotropic model}%
\ to the present computational model.

\subsection{Orthotropic material}%

\begin{figure}
[ptb]
\begin{center}
\includegraphics[width=0.65\textwidth]%
{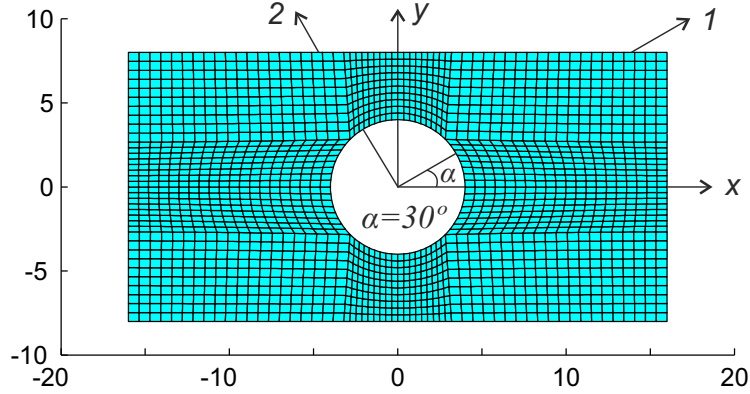}%
\caption{Rectangular plate with a concentric hole: reference configuration,
initial orientation ($\alpha=30{{}^o}$) of the preferred material directions
and finite element mesh. Dimensions of the plate: $l_{0}\times h_{0}%
=32\times16\operatorname{mm}^{2}$. Radius of the hole: $r_{0}%
=4\operatorname{mm}$.}%
\label{figure09.eps}%
\end{center}
\end{figure}
In this example we perform the analysis of an orthotropic visco-hyperelastic plate with a hole, see Figure \ref{figure09.eps}. The plate is loaded about the $x$--axis, i.e. $30º$ away the principal material $1$--axis. This example is the same as that given in Ref. \cite{LatMonCM2015}. In this case we used (bidimensional) $9/3$, \emph{u/p} mixed finite elements, see \cite{Bathe}. We have employed in the simulations the same time increments and time sequences as in Ref. \cite{LatMonCM2015}.

\begin{figure}
[ptb]
\begin{center}
\includegraphics[width=0.9\textwidth]%
{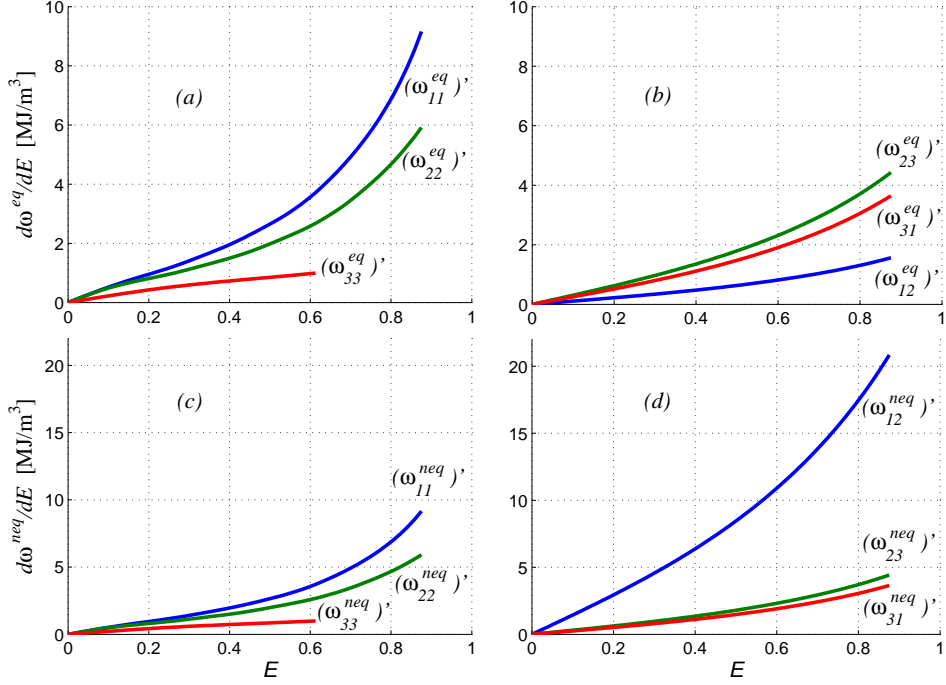}%
\caption{\emph{(a)} and \emph{(b)}: First derivative functions of the
components of the strain energy function $\mathcal{W}_{eq}$. \emph{(c)} and
\emph{(d)}: First derivative functions of the components of the strain energy
function $\mathcal{W}_{neq}$. Note that the only difference between
$\mathcal{W}_{eq}$ and $\mathcal{W}_{eq}$ is the component $\omega_{12}$. The
symmetries $\omega_{ij}^{\prime}(-E_{ij})=-\omega_{ij}^{\prime}(E_{ij})$ are
considered for all the functions shown in this figure.}%
\label{figure10.eps}%
\end{center}
\end{figure}

The deviatoric responses of the equilibrated and non-equilibrated parts of our
model are described by orthotropic spline-based strain energy functions of the
type ---c.f. Ref. \cite{LatMonCM2014}%
\begin{equation}
\mathcal{W}_{eq}(\mathbf{E}^{d})=%
{\displaystyle\sum\limits_{i=1}^{3}}
{\displaystyle\sum\limits_{j=1}^{3}}
\omega_{ij}^{eq}(E_{ij}^{d})
\end{equation}%
\begin{equation}
\mathcal{W}_{neq}(\mathbf{E}_{e}^{d})=%
{\displaystyle\sum\limits_{i=1}^{3}}
{\displaystyle\sum\limits_{j=1}^{3}}
\omega_{ij}^{neq}(E_{eij}^{d})
\end{equation}
whose first derivative functions are shown in Figure \ref{figure10.eps}.%

\begin{figure}
\begin{center}
\includegraphics[width=0.85\textwidth]%
{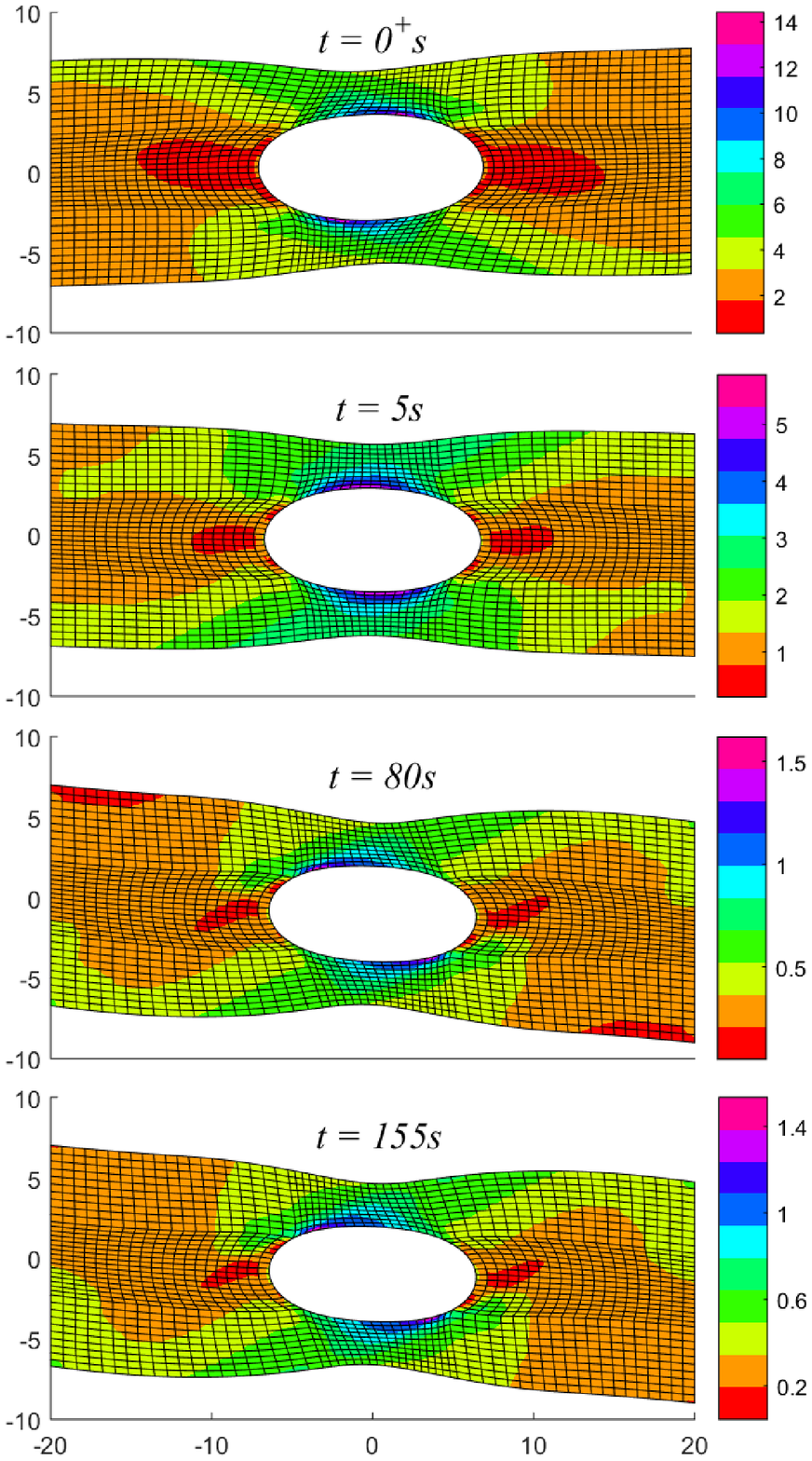}%
\caption{Relaxation process of the plate using the model based on the reverse
decomposition with an isotropic viscosity tensor.
Specifically, $\tau_{12}=3.78\operatorname{s}$. Deformed configurations and
distributions of $\left\Vert \mathbf{\sigma}^{d}\right\Vert $
($\operatorname{MPa}$) at instants $t=0^{+}\operatorname{s}$,
$t=5\operatorname{s}$, $t=80\operatorname{s}$ and $t=155\operatorname{s}$.
Unaveraged results at nodes.}%
\label{figure11.eps}%
\end{center}
\end{figure}

The preceding equilibrated and non-equilibrated stored energy functions were
used in the second simulation addressed within Example $3$ in Ref.
\cite{LatMonCM2015}. Therein, the prescribed value $\tau_{11}=10%
\operatorname{s}%
$ implied $\tau_{22}=10.23%
\operatorname{s}%
$, $\tau_{33}=23.86%
\operatorname{s}%
$, $\tau_{12}=3.78%
\operatorname{s}%
$, $\tau_{23}=17.82\operatorname{s}$ and $\tau_{31}=21.63\operatorname{s}$, thereby $\eta_{ij}^{d}=\eta^{d}$.
Even though the formulation employed in this case (based on the reversed
decomposition) is different from the formulation used in Ref.
\cite{LatMonCM2015} (based on the Sidoroff decomposition) we can see that very
similar results are obtained in both cases; compare Figure \ref{figure11.eps}
with Figure 13 of \cite{LatMonCM2015}.

\begin{figure}
[ptb]
\begin{center}
\includegraphics[width=0.85\textwidth]%
{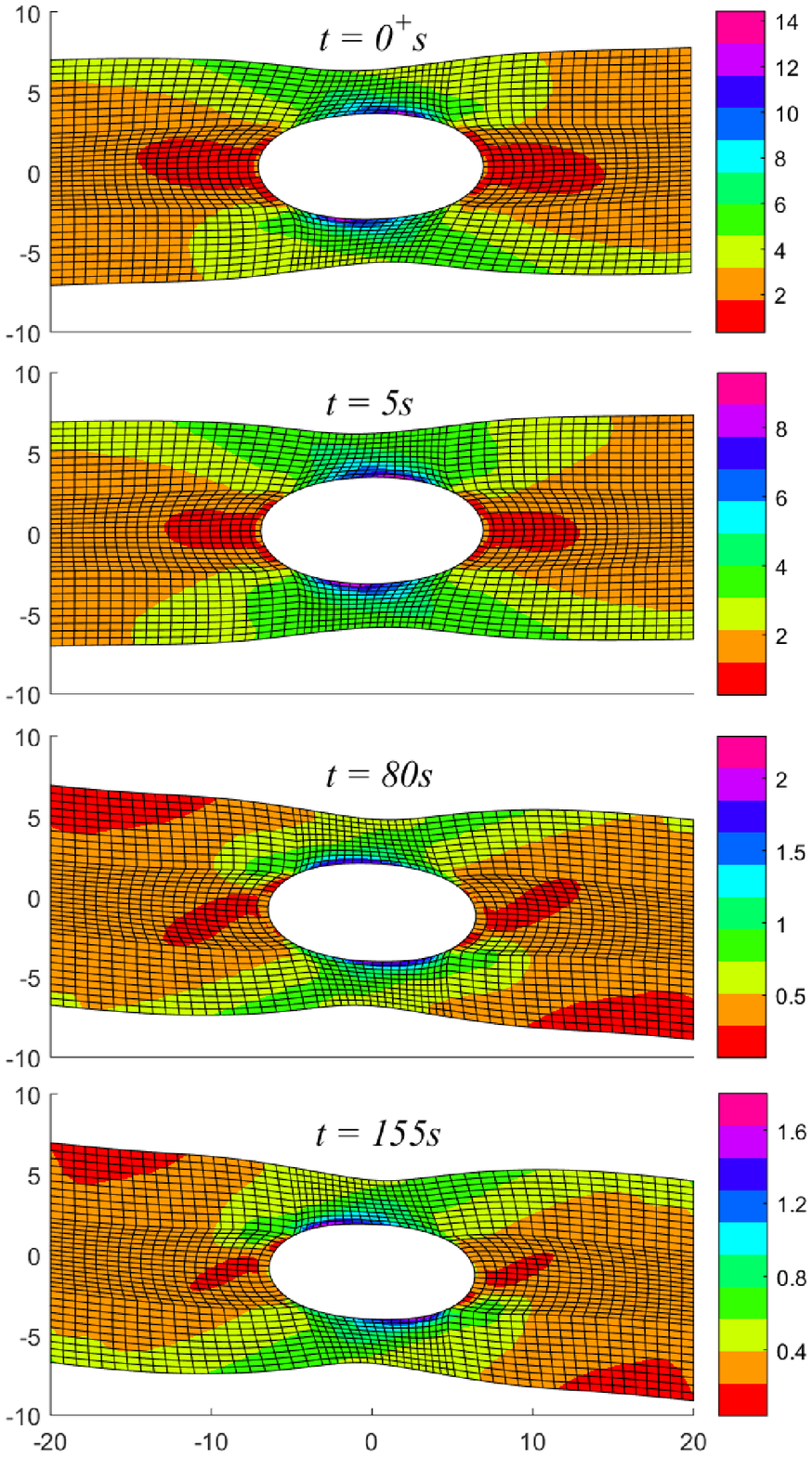}%
\caption{Relaxation process of the plate using the model based on the reversed
decomposition with an anisotropic viscosity tensor. Specifically, $\tau
_{12}=10\operatorname{s}$. Deformed configurations and distributions of
$\left\Vert \mathbf{\sigma}^{d}\right\Vert $ ($\operatorname{MPa}$) at
instants $t=0^{+}\operatorname{s}$, $t=5\operatorname{s}$,
$t=80\operatorname{s}$ and $t=155\operatorname{s}$. Unaveraged results at
nodes.}%
\label{figure12.eps}%
\end{center}
\end{figure}

In the second simulation addressed in the present example we modify the
relaxation time $\tau_{12}$ in order to show that with the present model we
have control over the relaxation process associated to the change of the
overall angular distortion from $\gamma_{xy}^{0}>0$ to $\gamma_{xy}^{\infty
}<0$. The values of the remaining relaxation times are preserved. The
relaxation time $\tau_{12}$ is increased up to $\tau_{12}=10%
\operatorname{s}%
$. As a result, note that the numerical calculations show a shear relaxation
process that is slower in Figure \ref{figure12.eps} than in Figure
\ref{figure11.eps}. Furthermore, we observe that a complete relaxation has
almost been achieved in Figure \ref{figure11.eps} at $t=155%
\operatorname{s}%
$, while the plate in Figure \ref{figure12.eps} is still relaxing at that
instant. Obviously, the other relaxation times could have been modified to
give other very different relaxation processes, always preserving the same
instantaneous and relaxed states. As a main consequence even though the present formulation is more complex, it is apparent that a
wider spectrum of material behaviors may be captured with the present model
than with that of Ref. \cite{LatMonCM2015}.

Convergence rates at representative steps using a symmetric and an unsymmetric
solver are shown in Table \ref{Tabla 2}. It can be seen that the use of a
symmetric solver results in only about one additional iteration per
step.
\begin{table}[h]%
\begin{tabular}
[c]{|llllll|}\hline
\begin{tabular}
[c]{@{}l}%
Time\\
\small ($\Delta t$)
\end{tabular}
&
\begin{tabular}
[c]{@{}l}%
Step\\
\small (Iteration)
\end{tabular}
&
\begin{tabular}
[c]{@{}l}%
Load norm\\
\small (Symmetric)
\end{tabular}
&
\begin{tabular}
[c]{@{}l}%
Load norm\\
\small (Unsymmetric)
\end{tabular}
&
\begin{tabular}
[c]{@{}l}%
Energy norm\\
\small (Symmetric)
\end{tabular}
&
\begin{tabular}
[c]{@{}l}%
Energy norm\\
\small (Unsymmetric)
\end{tabular}
\\\hline\hline
$2.5\operatorname{s}$ & 20(1) & 1.974E-01 & 1.974E-01 & 5.467E-03 &
5.466E-03\\
($0.125\operatorname{s}$) & 20(2) & 1.261E-02 & 1.281E-02 & 5.877E-07 & 2.872E-07\\
& 20(3) & 2.689E-05 & 8.682E-07 & 8.239E-11 & 1.975E-15\\
& 20(4) & 5.823E-07 &  & 3.830E-14 & \\
&  &  &  &  & \\
$20\operatorname{s}$ & 50(1) & 7.004E-01 & 7.004E-01 & 1.339E-01 &
1.339E-01\\
($1.5\operatorname{s}$) & 50(2) & 2.286E-01 & 2.331E-01 & 1.535E-04 & 1.460E-04\\
& 50(3) & 3.587E-04 & 2.225E-04 & 8.429E-09 & 1.223E-10\\
& 50(4) & 1.054E-05 & 4.378E-08 & 7.515E-12 & 3.893E-17\\
& 50(5) & 4.420E-07 &  & 1.249E-14 & \\\hline
\end{tabular}
\caption{Comparison of convergence rates using the reverse decomposition and
either a symmetric solver or an unsymmetric solver. Example of Figure \ref{figure12.eps}.}%
\label{Tabla 2}%
\end{table}

\section{Conclusions}

In this paper we present a phenomenological formulation and numerical
algorithm for anisotropic visco-hyperelasticity. The formulation is based on a
reverse multiplicative decomposition and on a split of the stored energy into
distinct anisotropic equilibrated and nonequilibrated addends. The formulation
is valid for large deviations from thermodynamic equilibrium. The procedure
may employ anisotropic stored energies and anisotropic viscosities. For the
orthotropic case, six relaxation experiments completely define the viscosities.
The procedure to obtain material parameters from experiments is also detailed.
The algorithm is formulated using logarithmic stress and strain measures in
order to facilitate the use of spline-based stored energies. The resulting
algorithmic tangent may be slightly nonsymmetric. However, in the analyzed
examples, the computational cost of using a symmetric tangent in terms of
iterations is small.

\section*{Acknowledgements}

Partial financial support for this work has been given by grant DPI2011-26635
from the Direcci\'{o}n General de Proyectos de Investigaci\'{o}n of the
Ministerio de Econom\'{\i}a y Competitividad of Spain.

\end{document}